\documentclass[lineno]{biometrika}

\nolinenumbers 

\usepackage{amsmath}
\usepackage{bbm}
\usepackage{booktabs} 
\usepackage{amsmath}  
\usepackage{amsfonts} 
\usepackage{tabularx}
\usepackage{subcaption}
\usepackage{array}
\usepackage{booktabs}
\usepackage{graphicx}

\usepackage{newtxtext}
\usepackage[subscriptcorrection]{newtxmath}

\usepackage{algorithm}
\usepackage{algpseudocode}
\usepackage{adjustbox}
\algrenewcommand\algorithmicrequire{\textbf{Input:}}
\algrenewcommand\algorithmicensure{\textbf{Output:}}

\newcommand{\1}[1]{\mathbbm{1}(#1)}

\usepackage{amsmath}
\usepackage[colorlinks=true,allcolors=black]{hyperref}
\usepackage[noabbrev,capitalise]{cleveref}
\creflabelformat{equation}{#2\textup{#1}#3}

\crefname{algo}{Algorithm}{Algorithms}       
\Crefname{algo}{Algorithm}{Algorithms} 
\definecolor{OliveGreen}{rgb}{0,0.6,0}
\definecolor{magenta}{rgb}{1,0,1}

\newcommand{\branch}[4]{
\left\{
	\begin{array}{ll}
		#1  & \mbox{if } #2 \\
		#3 & \mbox{if } #4
	\end{array}
\right.
}

\begin{document}

\jname{} 

\markboth{C.\! T.\ COVINGTON \& J.\! W.\ MILLER}{PITOS goodness-of-fit tests}

\title{A powerful goodness-of-fit test using the probability integral transform of order statistics}

\author{Christian T.\ Covington}
\affil{Department of Biostatistics, Harvard T.H. Chan School of Public Health,\\ 677 Huntington Avenue, Boston, MA 02115, U.S.A.\email{ccovington@g.harvard.edu}}

\author{Jeffrey W.\ Miller}
\affil{Department of Biostatistics, Harvard T.H. Chan School of Public Health,\\ 677 Huntington Avenue, Boston, MA 02115, U.S.A. \email{jwmiller@hsph.harvard.edu}}

\maketitle
\thispagestyle{empty}

\begin{abstract}
  Goodness-of-fit (GoF) tests are a fundamental component of statistical practice, essential for 
  checking model assumptions and testing scientific hypotheses. 
  Despite their widespread use, popular GoF tests exhibit surprisingly low statistical power against 
  substantial departures from the null hypothesis. 
  To address this, we introduce PITOS, a novel GoF test based on applying the probability integral transform (PIT) to the $j$th order statistic (OS) given the $i$th order statistic for selected pairs $i,j$.
  Under the null, for any pair $i,j$, this yields a $\mathrm{Uniform}(0,1)$ random variable, which we map to a p-value via $u\mapsto 2\min(u, 1-u)$. 
  We compute these p-values for a structured collection of pairs $i,j$ generated via a discretized transformed Halton sequence, and aggregate them using the Cauchy combination technique to obtain the PITOS p-value.
  Our method maintains approximately valid Type I error control, 
  has an efficient $O(n \log n)$ runtime, and can be used with any null distribution via the Rosenblatt transform.
  In empirical demonstrations, we find that PITOS has much higher power than popular GoF tests on distributions characterized by local departures from the null, while maintaining competitive power across all distributions tested.
\end{abstract}


\section{Introduction}
\label{sec:introduction}

Goodness-of-fit (GoF) testing is a key part of the statistical toolkit, widely used for validating modeling assumptions and testing scientific hypotheses.
By design, GoF tests typically assess the fit of observed data to a null distribution without requiring a pre-specified alternative. 
This generality makes GoF tests very flexible, but also means that they are often applied with minimal regard to the types of deviations from the null that might occur. 
As such, ideally a GoF test would have high power to detect a wide range of deviations from the null \citep{dagostino1986goftechniques, lehmann2005testing}. 

\begin{figure}
    \centering
    \includegraphics[width=\textwidth]{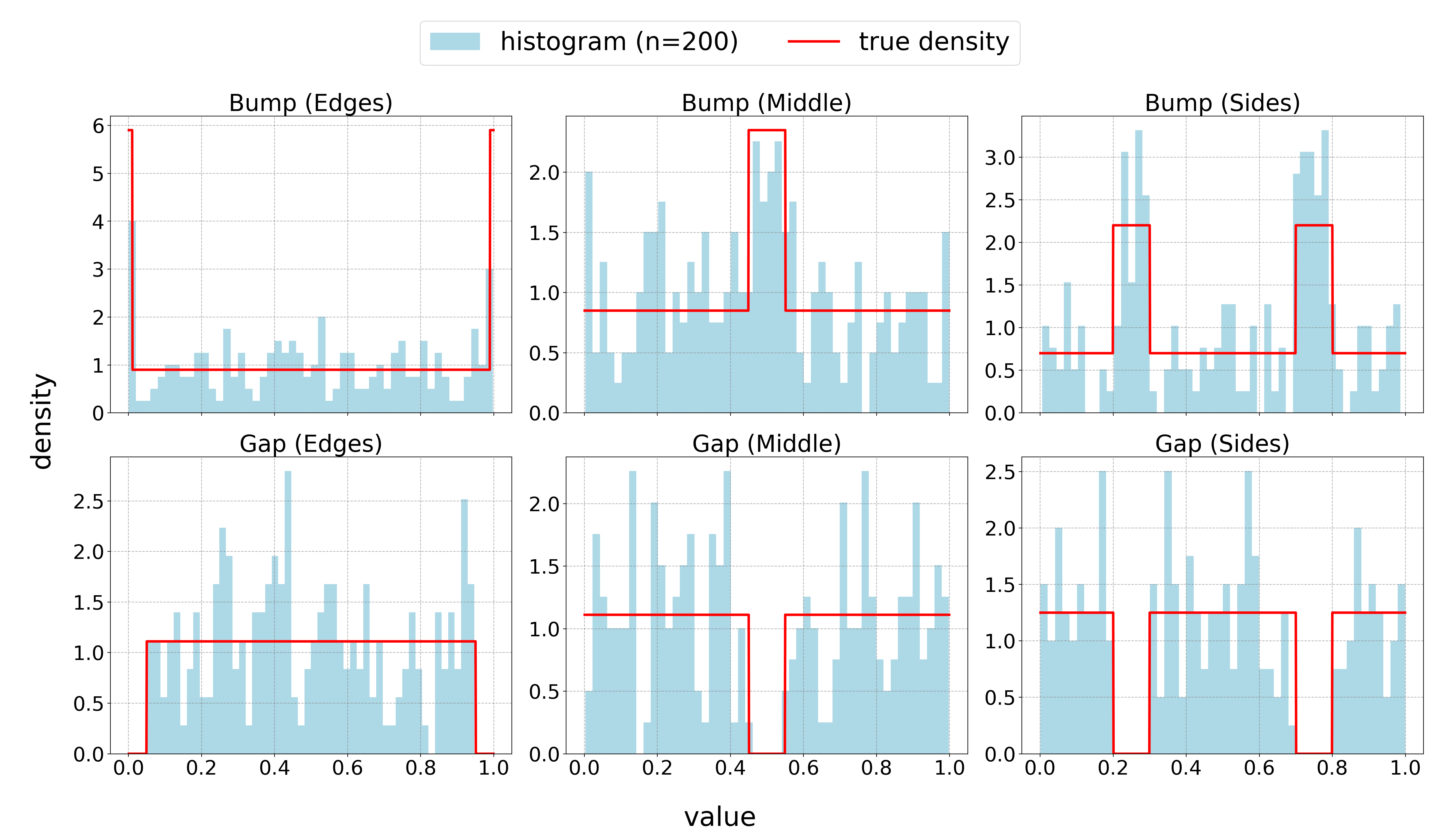}
    \caption{Examples of data distributions on which popular GoF tests have low power. For each, the true density is shown in red, and a histogram of one simulated data set is shown in blue.}
    \label{fig:stark_failures}
\end{figure}

However, the most popular GoF tests often have surprisingly low statistical power, even on distributions that are conspicuously different from the null.
To illustrate, we compute the power of four popular GoF tests on data from the six distributions shown in \cref{fig:stark_failures}, taking $\mathrm{Uniform}(0,1)$ to be the null hypothesis.  Specifically, we consider the Anderson--Darling (AD), Neyman--Barton (NB), Kolmogorov--Smirnov (KS), and Cram\'{e}r--von Mises (CvM) tests; see \cref{sec:standard_gof_tests}.

In \cref{tab:power_stark_failures}, we see that each of these tests exhibits low power on most of these distributions.
In particular, KS has very low power in the tails, AD has very low power around the median, NB has very low power around the first and third quartiles, and CvM has very low power across the board.
This poor performance is not because these departures from the null are intrinsically hard to detect. Indeed, power very close to 1 is obtained by an oracle benchmark based on the likelihood ratio test (LRT) using the true distribution as the alternative hypothesis (\cref{tab:power_stark_failures}); see \cref{sec:standard_gof_tests} for details.

\begin{table}
    \centering
    
    \resizebox{\textwidth}{!}{%
    \begin{tabular}{|l|l|l|l|l|l|l|}
        \hline
        Test                & Bump (Edges) & Gap (Edges)  & Bump (Middle) & Gap (Middle) & Bump (Sides) & Gap (Sides) \\ \hline
        Anderson--Darling   & 0.843       & 0.529         & 0.387       & \textcolor{red}{0.205}  & 0.392                  & 0.256 \\ \hline
        Neyman--Barton      & 0.765       & 0.810         & 0.540 & 0.327                    & \textcolor{red}{0.114} & \textcolor{red}{0.152} \\ \hline
        Kolmogorov--Smirnov & \textcolor{red}{0.189} & \textcolor{red}{0.114} & 0.593         & 0.376 & 0.516 & 0.387 \\ \hline
        Cram\'{e}r--von Mises & \textcolor{red}{0.216} & \textcolor{red}{0.152} & 0.428         & \textcolor{red}{0.197} & 0.257  & \textcolor{red}{0.154} \\ \hline
        LRT (Oracle)  & 1.0 & 1.0 & 1.0 & 1.0 & 1.0 & 1.0 \\ \hline
    \end{tabular}%
    }
    \caption{Power of commonly used GoF tests on data from the distributions in~\cref{fig:stark_failures}. We approximate the power at the $\alpha = 0.05$ level by averaging over $100{,}000$ simulated data sets with a sample size of $n=200$. Red text indicates power $\leq$ 0.25.}
    \label{tab:power_stark_failures}
\end{table}

We introduce a novel GoF test, referred to as the  \textit{Probability Integral Transform of Order Statistics} (PITOS), that has high power in many cases where these popular tests fail and has competitive power across all cases that we have explored.
PITOS takes a sequence of pairs of indices $i,j\in\{1,\ldots,n\}$, uses the conditional distribution of the order statistics $X_{(j)}\mid X_{(i)}$ under the null to calculate a p-value $p_{i j}$ for each pair $(i,j)$, and combines these p-values via Cauchy combination to get a single combined p-value. This approach admits a large class of tests via the choice of pairs $(i,j)$, so we propose a default which generates pairs of indices via a discretized transformed Halton sequence. We show that PITOS maintains approximately valid Type I error control and has an $O(n \log n)$ runtime. We perform simulation studies comparing the power of PITOS and other GoF tests on a range of distributions.

\section{Methodology}
\label{sec:methodology}

In this section, we introduce our proposed goodness-of-fit test, the \textit{Probability Integral Transform of Order Statistics} (PITOS).
Given independent and identically distributed (i.i.d.) data $X_1,\ldots,X_n$, 
we are interested in testing whether we can reject a point null hypothesis, which we take to be that $X_1,\ldots,X_n\sim\mathrm{Uniform}(0,1)$.  
The choice of $\mathrm{Uniform}(0,1)$ is made without loss of generality since for any null with a continuous cumulative distribution function (CDF) $F$, if $Y_i\sim F$ then $F(Y_i) \sim \mathrm{Uniform}(0,1)$; thus, we can first map the data through $F$ and then apply a test based on a uniform null.
More generally, any random vector $(Y_1,\ldots,Y_n)$ with a known joint distribution can be mapped to a vector of i.i.d.\ $\textrm{Uniform}(0,1)$ random variables via the generalized Rosenblatt transform \citep{brockwell2007universal}; see~\cref{sec:generalized_rosenblatt} for details.

\subsection{PITOS goodness-of-fit test}

Let $X_1,\ldots,X_n$ i.i.d.\ $\sim\mathrm{Uniform}(0,1)$.
The order statistics $X_{(1)},\ldots,X_{(n)}$ are defined by arranging $X_1,\ldots,X_n$ in non-decreasing order.
It is well-known that the CDF of $X_{(j)}$ is 
$F_{X_{(j)}}(y) = G(y, j, n-j+1)$, 
where $G(x,a,b)$ is the CDF of the $\mathrm{Beta}(a,b)$ distribution evaluated at $x$, also known as the regularized incomplete beta function, $I_x(a,b)$.
Furthermore, for any distinct $i,j\in\{1,\ldots,n\}$, the CDF of the conditional distribution of $X_{(j)}$ given $X_{(i)}$ is 
\begin{align}
    \label{eq:order_stat_cond_cdf}
    F_{X_{(j)} \vert X_{(i)} = x}(y) = \branch{G((y - x)/(1-x), j-i, n-j+1)~~}{i < j,}{G(y/x, j, i - j)}{i > j;} 
\end{align}
see Theorems 2.4.1 and 2.4.2 of~\citet{arnold2008first} for reference.
We apply the probability integral transform to these marginal and conditional distributions. Specifically, defining
\begin{align*}
    h_{i j}(x,y) = \branch{F_{X_{(j)}}(y)}{i = j,}{F_{X_{(j)} \vert X_{(i)} = x}(y)}{i\neq j,}
\end{align*}
and letting $U_{i j} := h_{i j}(X_{(i)}, X_{(j)})$, it follows that $U_{i j} \sim \mathrm{Uniform}(0,1)$.
Values of $U_{i j}$ close to $0$ or $1$ indicate that $X_{(j)}$ is smaller or larger than expected, given $X_{(i)}$ when $i\neq j$ or marginally when $i=j$.
Thus, we construct a p-value $p_{i j} := 2\min(U_{i j}, 1 - U_{i j})$, which is small whenever $X_{(j)}$ is unusually small or large, relative to $X_{(i)}$.

Now, suppose $\mathcal{I} = ((i_1,j_1),\ldots,(i_m,j_m))$ is a sequence of $m$ pairs $(i,j)$. 
We propose to compute the p-values $p_{i j}$ as defined above for all $(i,j)$ in $\mathcal{I}$ and then aggregate them using the Cauchy combination technique \citep{liu2020cauchy} to obtain a combined p-value,
\begin{align}
\label{eq:cauchy_combo}
    p := 1 - F_{\mathrm{Cauchy}}\bigg(\frac{1}{m}\sum_{(i,j)\in\mathcal{I}} F_{\mathrm{Cauchy}}^{-1}(1 - p_{i j})\bigg)
\end{align}
where $F_{\mathrm{Cauchy}}$ is the CDF of the $\mathrm{Cauchy}(0,1)$ distribution.
Finally, to adjust for the fact that the Cauchy combination does not exactly control Type I error due to the dependence among the $p_{i j}$ values, we use a simple multiplicative correction, $p^* := \min(1, 1.15 p)$. Empirically, we find that $p^*$ approximately controls Type I error at any level $\alpha\in(0,1)$, while attaining Type I error rates close to $\alpha$ when $\alpha \leq 0.05$; see \cref{supp:fig:PITOS_nulls}.

In \cref{sec:algorithms}, we provide a step-by-step algorithm for computing the PITOS p-value $p^*$.

\subsection{Choosing a sequence of pairs}

The PITOS test requires specification of a sequence of pairs $\mathcal{I}$, and the choice of $\mathcal{I}$ affects the power to detect different alternative hypotheses.
Based on simulation studies, we find that any given pair $(i,j)$ may be powerful for detecting certain alternatives, but not others.  Furthermore, the power of $(i,j)$ is not equal to the power of $(j,i)$, in general.
Roughly speaking, the power of a pair $(i,j)$ is higher for departures from the null between the $i/n$ and $j/n$ quantiles.
In other words, $(i,j)$ tends to have good power to detect an alternative with density higher or lower than $1$ in the interval from $i/n$ to $j/n$.
In particular, values of $j/n$ close to $0$ or $1$ tend to be useful for detecting departures in the left or right tail, respectively.

Thus, to obtain good power across a wide range of alternatives, we aim to choose a sequence of points $(i,j)$ such that $(i/n,j/n)$ is well-distributed throughout the unit square $[0,1]^2$, with heavier coverage near the boundaries, since detecting tail deviations is of particular importance in many practical applications.
It is also desirable to include additional points $(i,i)$ on the diagonal, since the marginal order statistics tend to have good power for distributions close to $\mathrm{Uniform}(0,1)$.
Note that a given point $(i,j)$ may appear more than once in $\mathcal{I}$, and this has the effect of upweighting the contribution of that pair in the Cauchy combination of p-values in \cref{eq:cauchy_combo}.

Based on these considerations, we propose to generate $\mathcal{I}$ as follows.
For $k = 1,\ldots,m-n$, generate $(u_k,v_k)$ from the two-dimensional Halton sequence on $[0,1]^2$ with bases $2$ and $3$ \citep{halton1964algorithm}. Transform to $x_k = F_{\mathrm{Beta}(0.7,0.7)}^{-1}(u_k)$ and $y_k = F_{\mathrm{Beta}(0.7,0.7)}^{-1}(v_k)$, that is, map $u_k$ and $v_k$ through the inverse CDF of the $\mathrm{Beta}(0.7,0.7)$ distribution, and set $i_k = \lceil n x_k \rceil$ and $j_k = \lceil n y_k \rceil$.
Finally, set $i_{m-n+r} = r$ and $j_{m-n+r} = r$ for $r = 1,\ldots,n$, and define $\mathcal{I} = ((i_1,j_1),\ldots,(i_m,j_m))$.
To keep the algorithm as computationally efficient as possible, it makes sense to choose $m = O(n\log n)$ since computing the order statistics $X_{(1)},\ldots,X_{(n)}$ already takes $O(n \log n)$ time.
We recommend using $m = \lceil 10\, n\log n\rceil + n$ since we find that this provides good empirical performance.
See \cref{sec:algorithms} for a step-by-step algorithm.

Alternatively, one could use random samples of $(u_k,v_k)\sim\mathrm{Uniform}([0,1]^2)$, but the Halton sequence has the advantage of producing points that are more evenly distributed than a random sequence, and has the added benefit of being deterministic.
The procedure above could also be customized by replacing the $\mathrm{Beta}(0.7,0.7)$ CDF with any other CDF on $[0,1]$, in order to try to increase power for a specific collection of alternative distributions of interest.
\section{Experiments}
\label{sec:experiments}

In this section, we compare the performance of PITOS (our method) and several benchmark GoF tests in simulation studies.
As benchmarks, we compare with the Anderson--Darling (AD), Neyman--Barton (NB), Kolmogorov--Smirnov (KS), and Cram\'{e}r--von Mises (CvM) tests; see \cref{sec:standard_gof_tests} for details.
To quantify the highest power possible, we also compare with an oracle benchmark based on the likelihood ratio test (LRT) using the true distribution as a point alternative; see \cref{sec:standard_gof_tests}.
This represents the optimal power since the LRT is the most powerful test for a specific null/alternative pair, by the Neyman--Pearson lemma~\citep{neyman1933ix}, but this test cannot be used in practice because it requires knowing the true distribution to compute the test statistic.

\begin{figure}
    \centering
    \includegraphics[width=\textwidth]{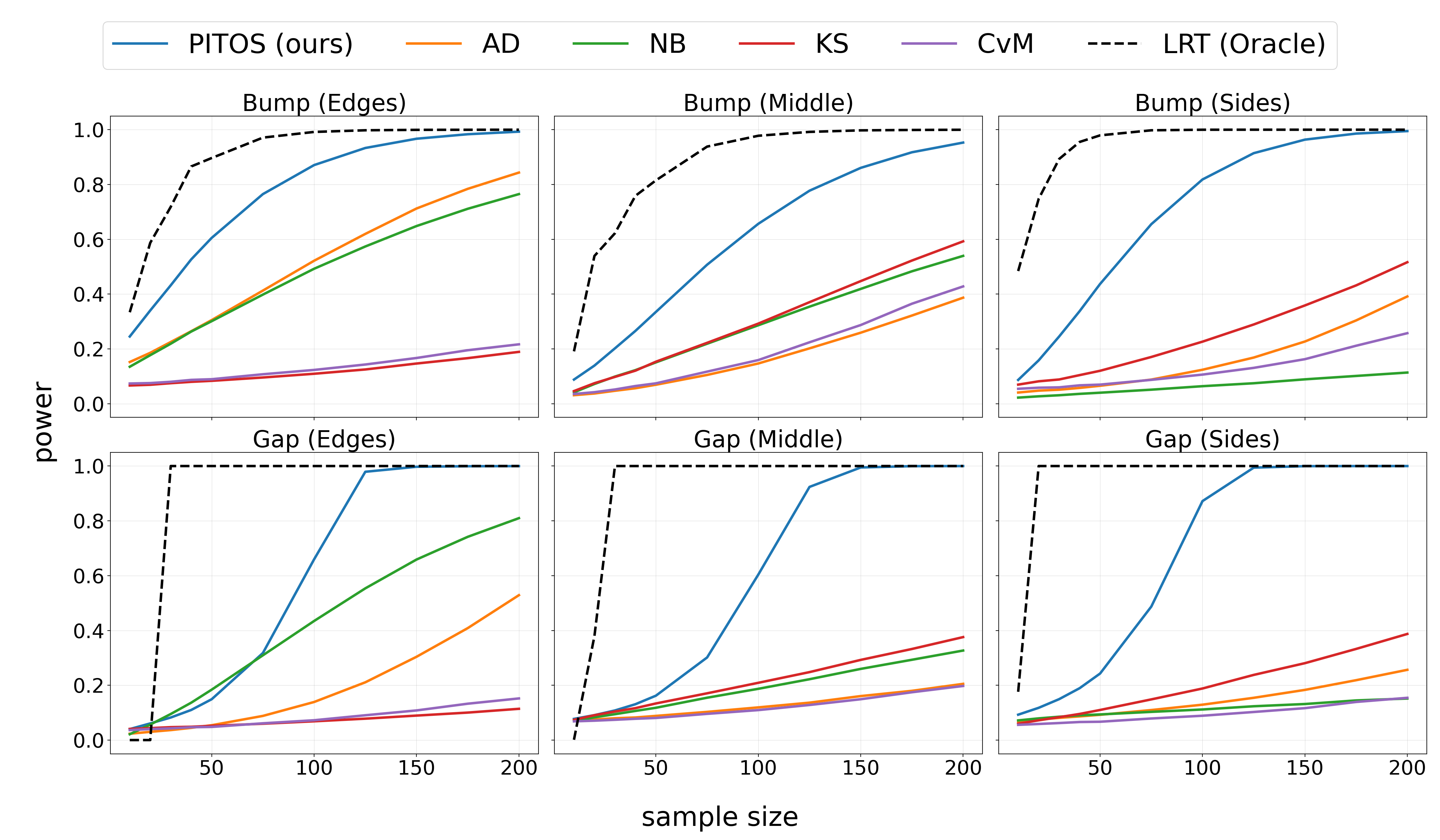}

    \caption{Power of each method versus sample size $n$ for the distributions in~\cref{fig:stark_failures}.}
    \label{fig:power_stark_failures}
\end{figure}

First, we consider the power of each test on the distributions from \cref{fig:stark_failures}, which represent motivating examples where the commonly used tests generally have low power.
For a given sample size $n$, we compute the power to reject the null at the $\alpha = 0.05$ level by generating 100{,}000 simulated i.i.d.\ data sets from the true distribution and calculating the proportion of times that the p-value is less than $\alpha = 0.05$.
\cref{fig:power_stark_failures} shows the power of each test as a function of $n$. 
We see that PITOS outperforms the other tests by a wide margin on all of these distributions.
The value of these curves at $n = 200$ corresponds to the values of the power shown in \cref{tab:power_stark_failures}.

\begin{figure}
    \centering
    \includegraphics[width=\textwidth]{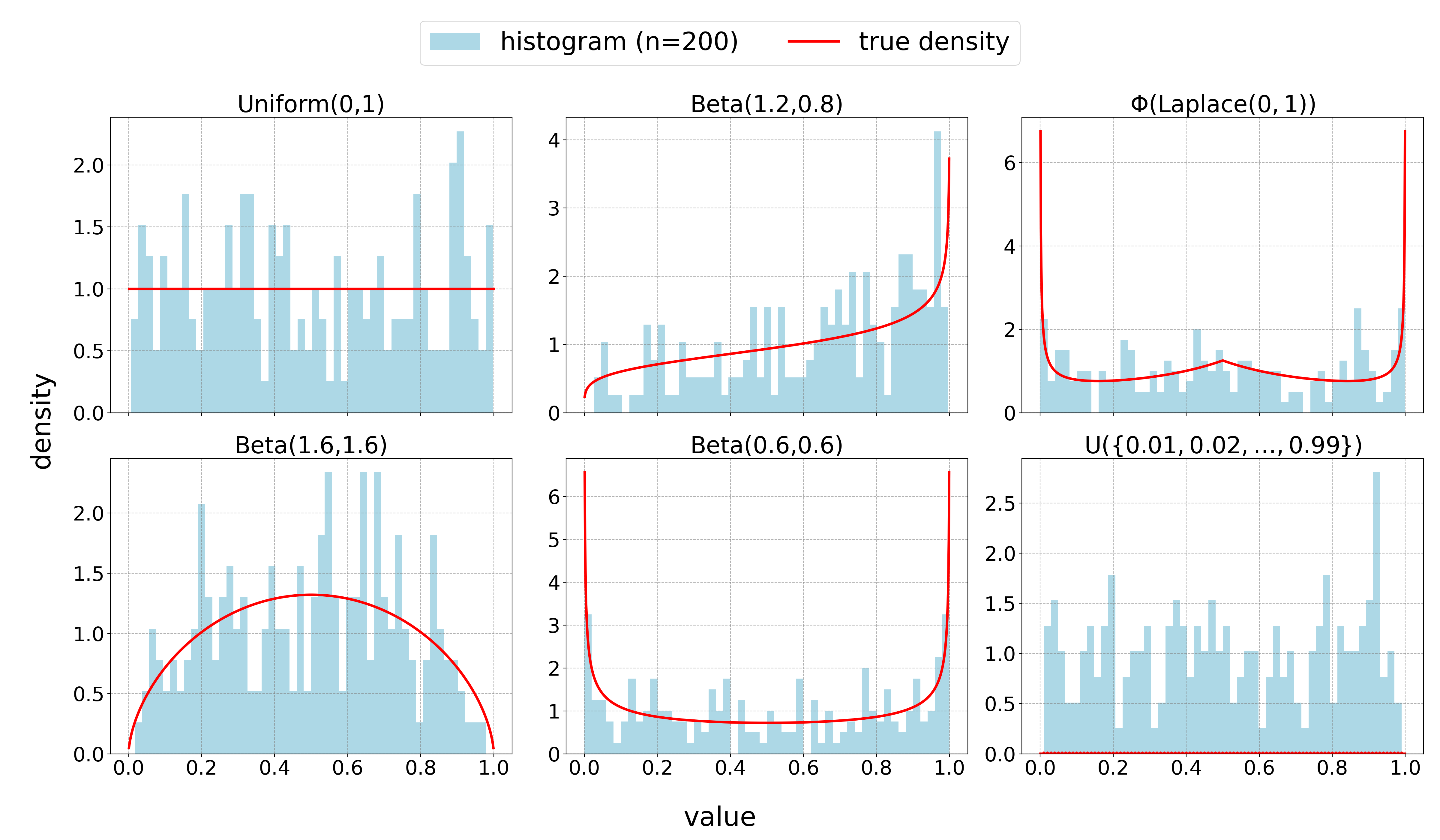}
    \\~\\
    \includegraphics[width=\textwidth]{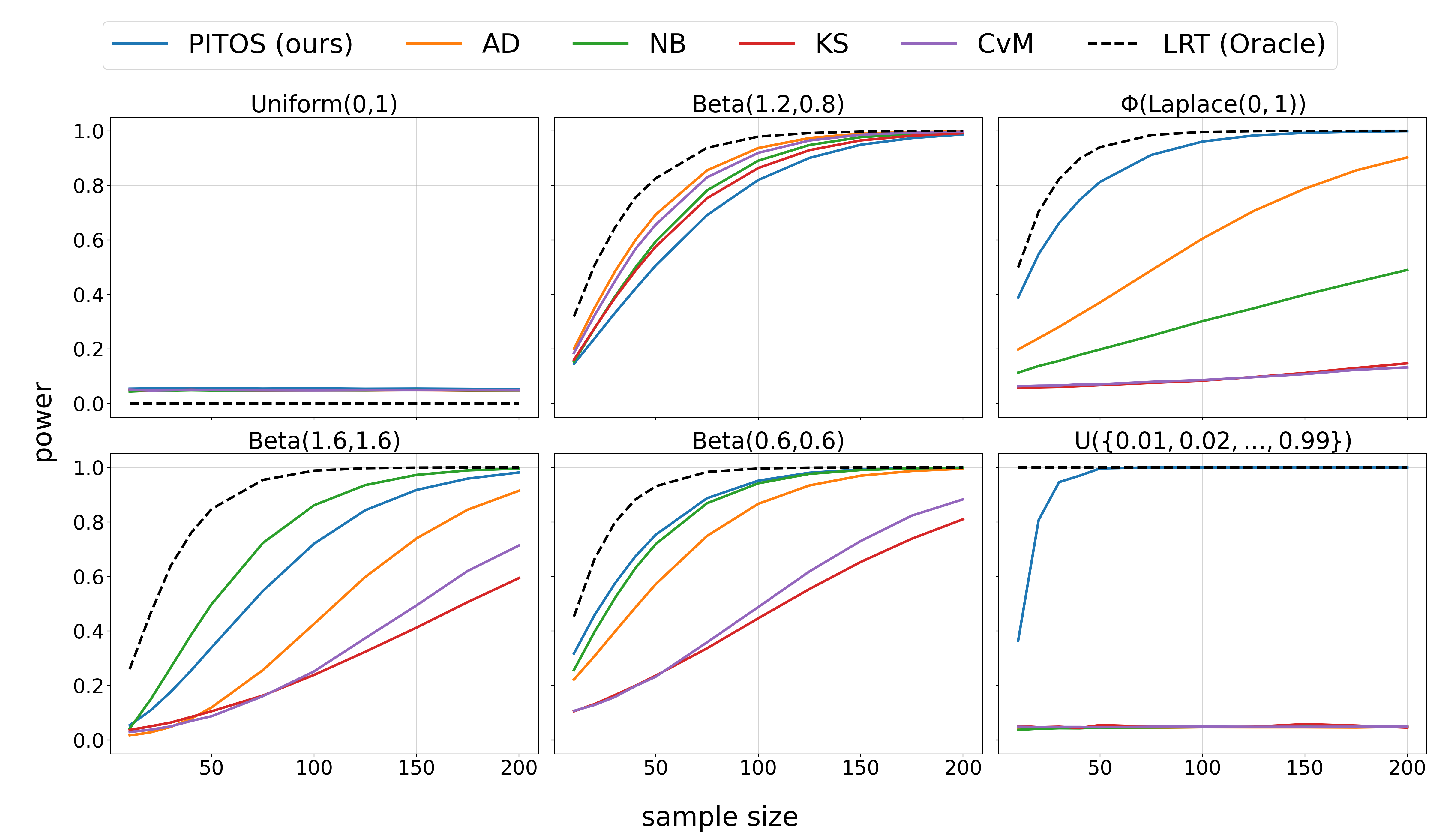}
    \caption{Power of GoF tests across various data distributions.}
    \label{fig:regular}
\end{figure}

Next, we consider the set of six distributions shown in \cref{fig:regular} (top), chosen to represent a range of behaviors.
\cref{fig:regular} (bottom) shows the power at level $\alpha = 0.05$ versus $n$ for each distribution, again using 100{,}000 simulated i.i.d.\ data sets for each $n$.
On $\mathrm{Uniform}(0,1)$ data---that is, when the null hypothesis is true---all methods have power close to $0.05$, indicating that they correctly control Type I error. (The LRT power is $0$ here because the test statistic is identically constant.)
The $\mathrm{Beta}(1.2,0.8)$ distribution is moderately close to uniform, and here we see that PITOS has the lowest power, although the differences between methods are relatively small.
Meanwhile, PITOS is best or second-best on $\mathrm{Beta}(0.6,0.6)$ and $\mathrm{Beta}(1.6,1.6)$, which represent cases with somewhat higher or lower density, respectively, in the tails.
PITOS excels when there are outliers or heavy tails, such as in the case of $\Phi(\mathrm{Laplace}(0,1))$, which is the distribution of $\Phi(Y)$ where $Y\sim\mathrm{Laplace}(0,1)$ and $\Phi(\cdot)$ is the standard normal CDF.
Furthermore, PITOS tends to have high power to detect discreteness, such as in the case of $\mathrm{Uniform}(\{0.01,0.02,\ldots,0.99\})$, the discrete uniform distribution on $\{0.01,0.02,\ldots,0.99\}$. The other methods have power essentially at $\alpha = 0.05$ on this discrete example, meaning that they completely fail to detect this departure from the null of $\mathrm{Uniform}(0,1)$.

\begin{figure}
    \centering
    \includegraphics[width=0.48\textwidth]{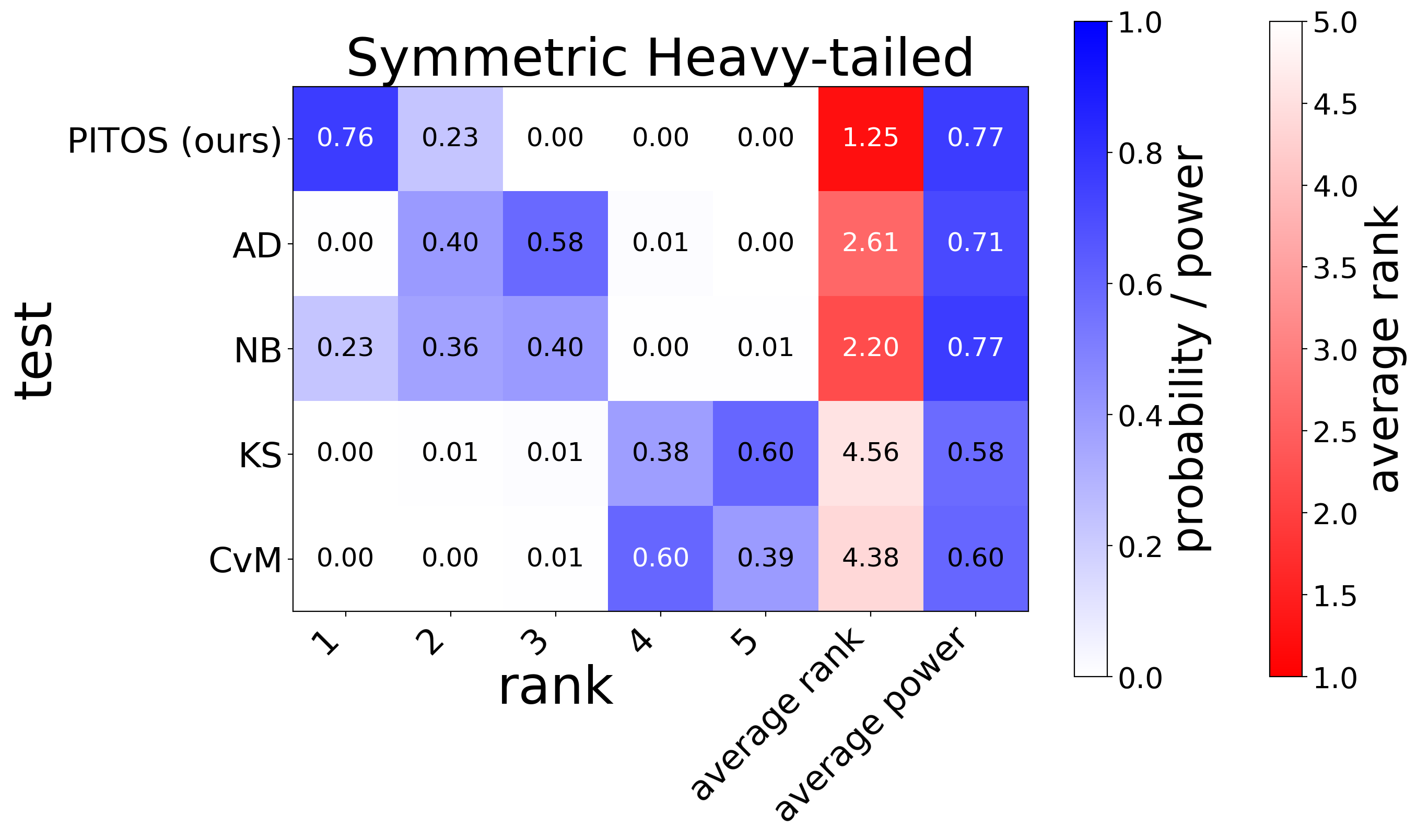}
    \hfill
    \includegraphics[width=0.48\textwidth]{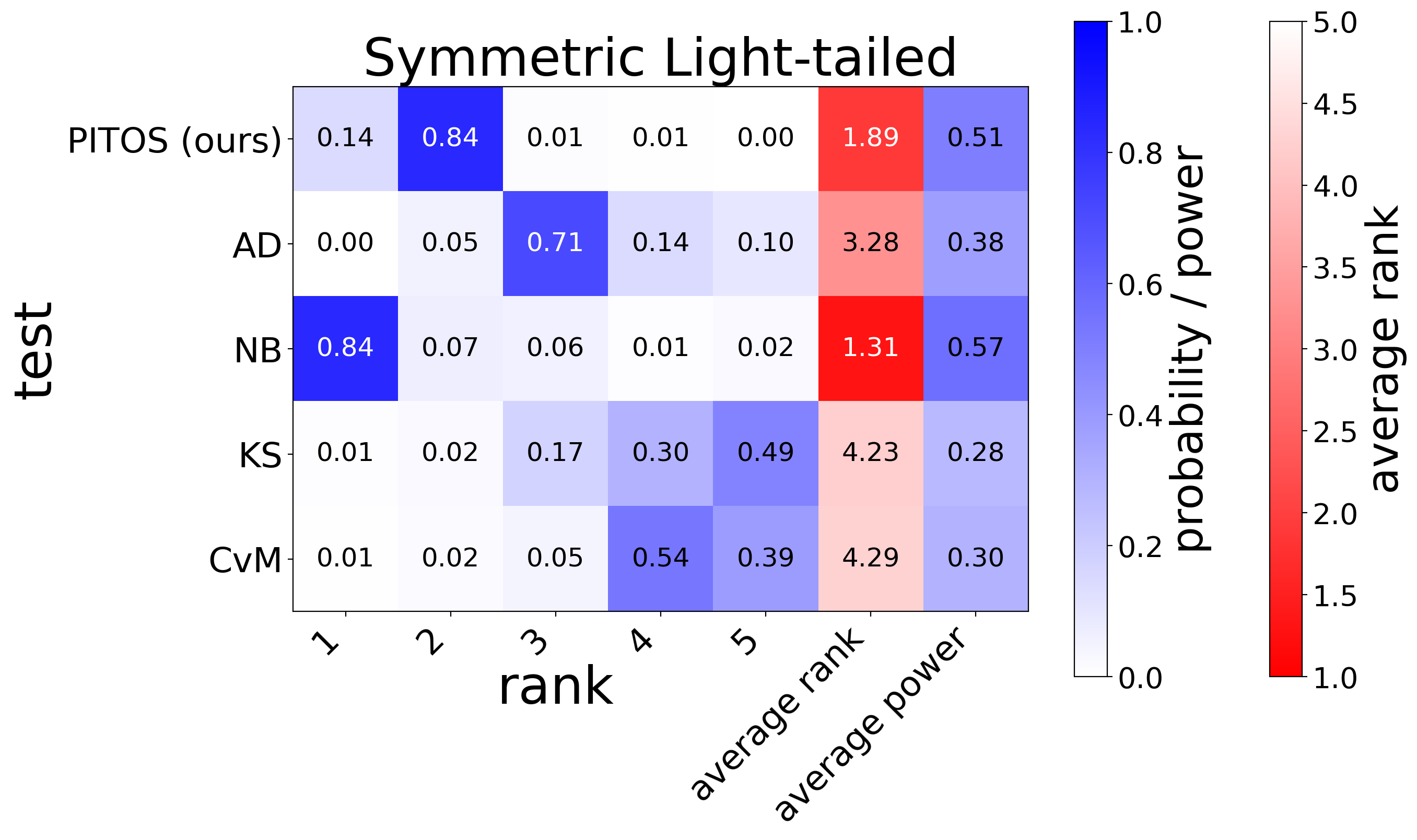}
    \hfill
    \includegraphics[width=0.48\textwidth]{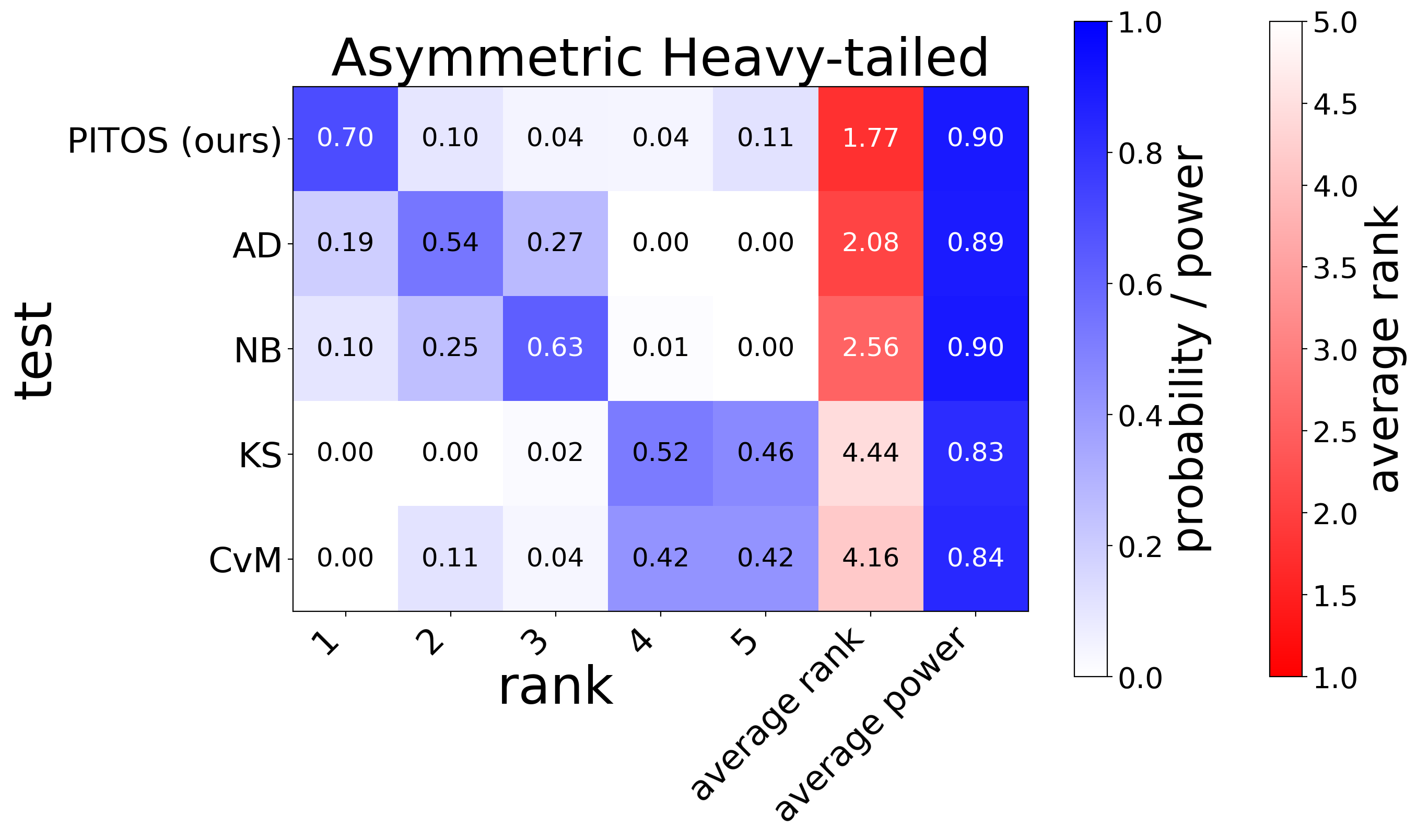}
    \hfill
    \includegraphics[width=0.48\textwidth]{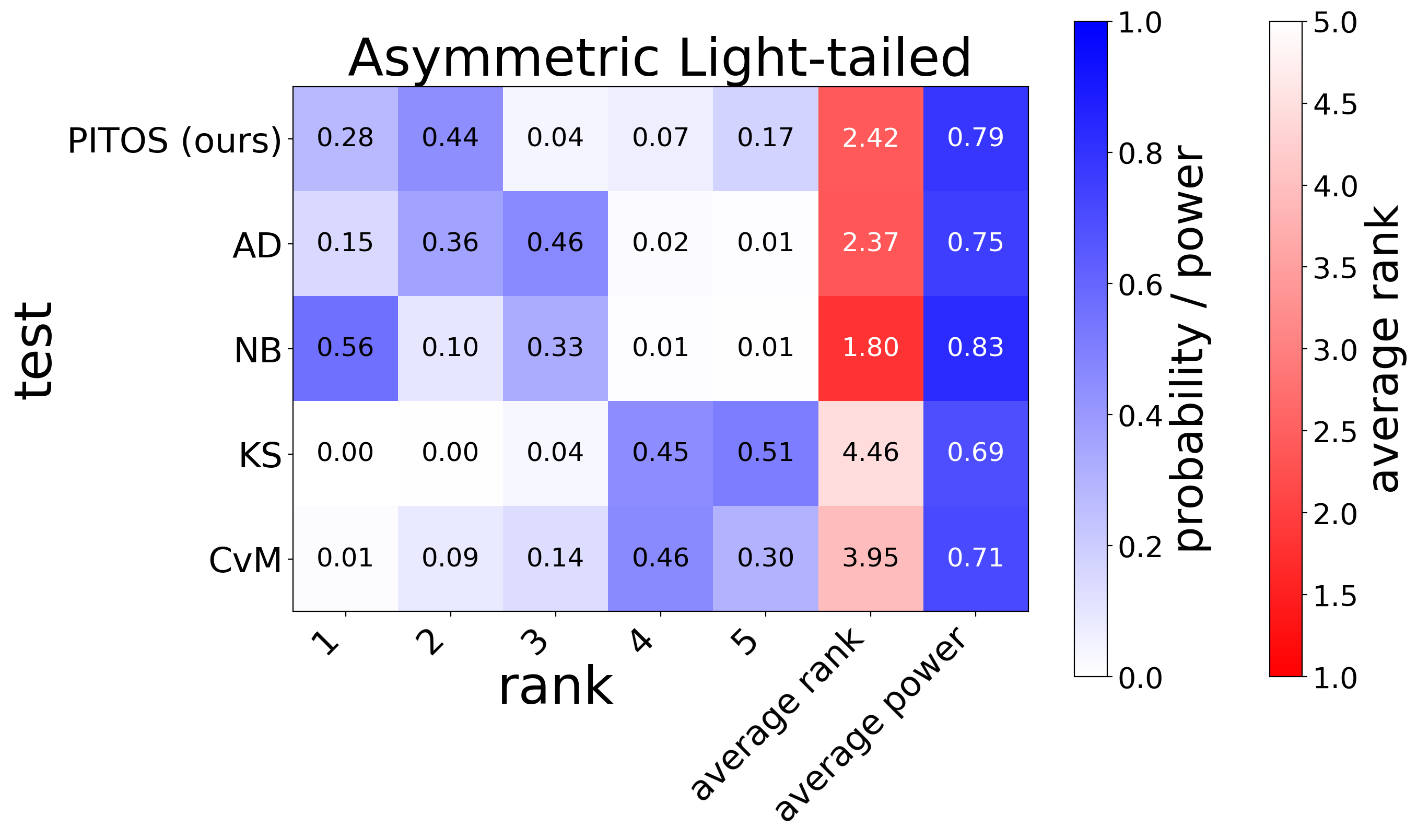}
    \hfill
    \includegraphics[width=0.48\textwidth]{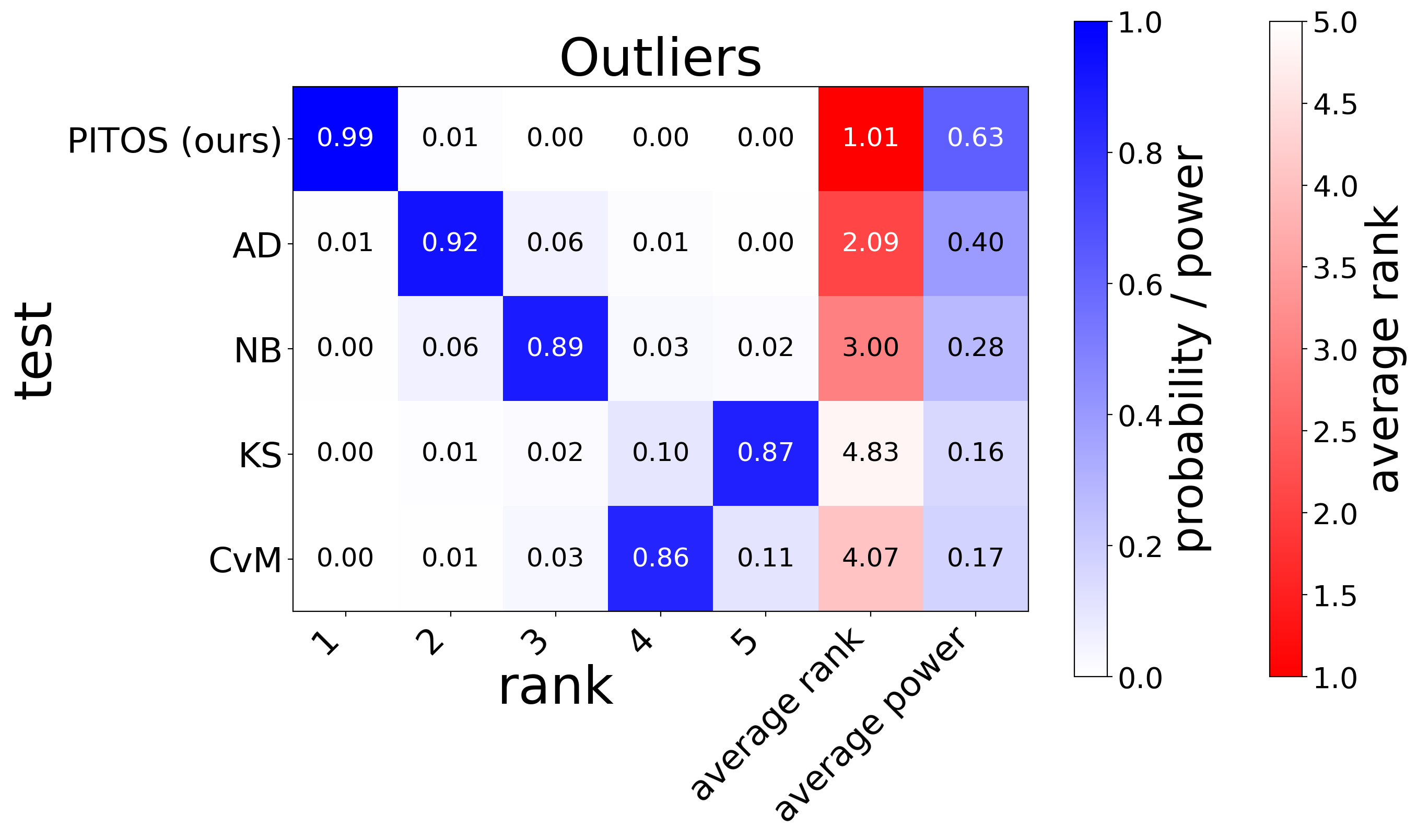}
    \hfill
    \includegraphics[width=0.48\textwidth]{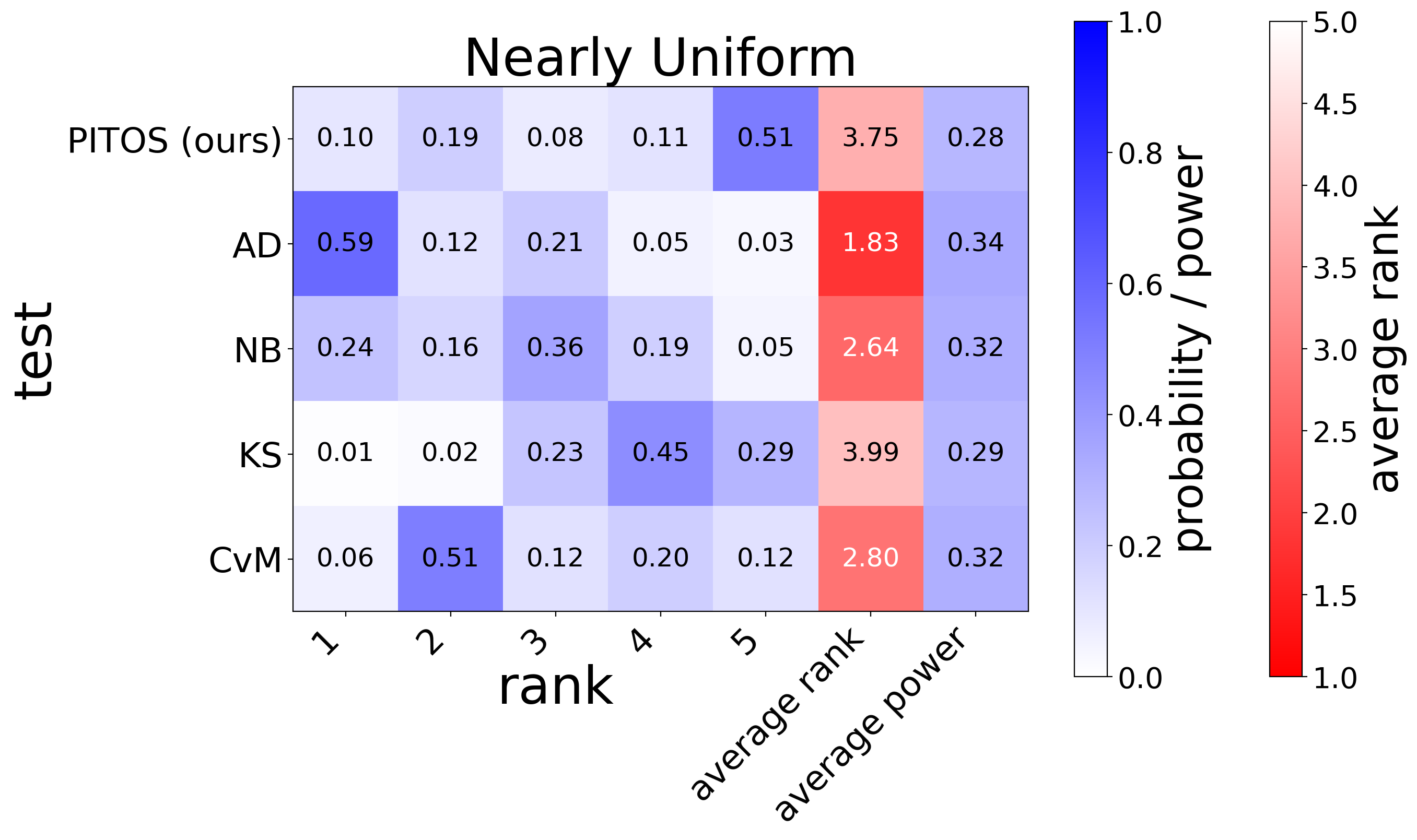}
    \hfill
    \includegraphics[width=0.48\textwidth]{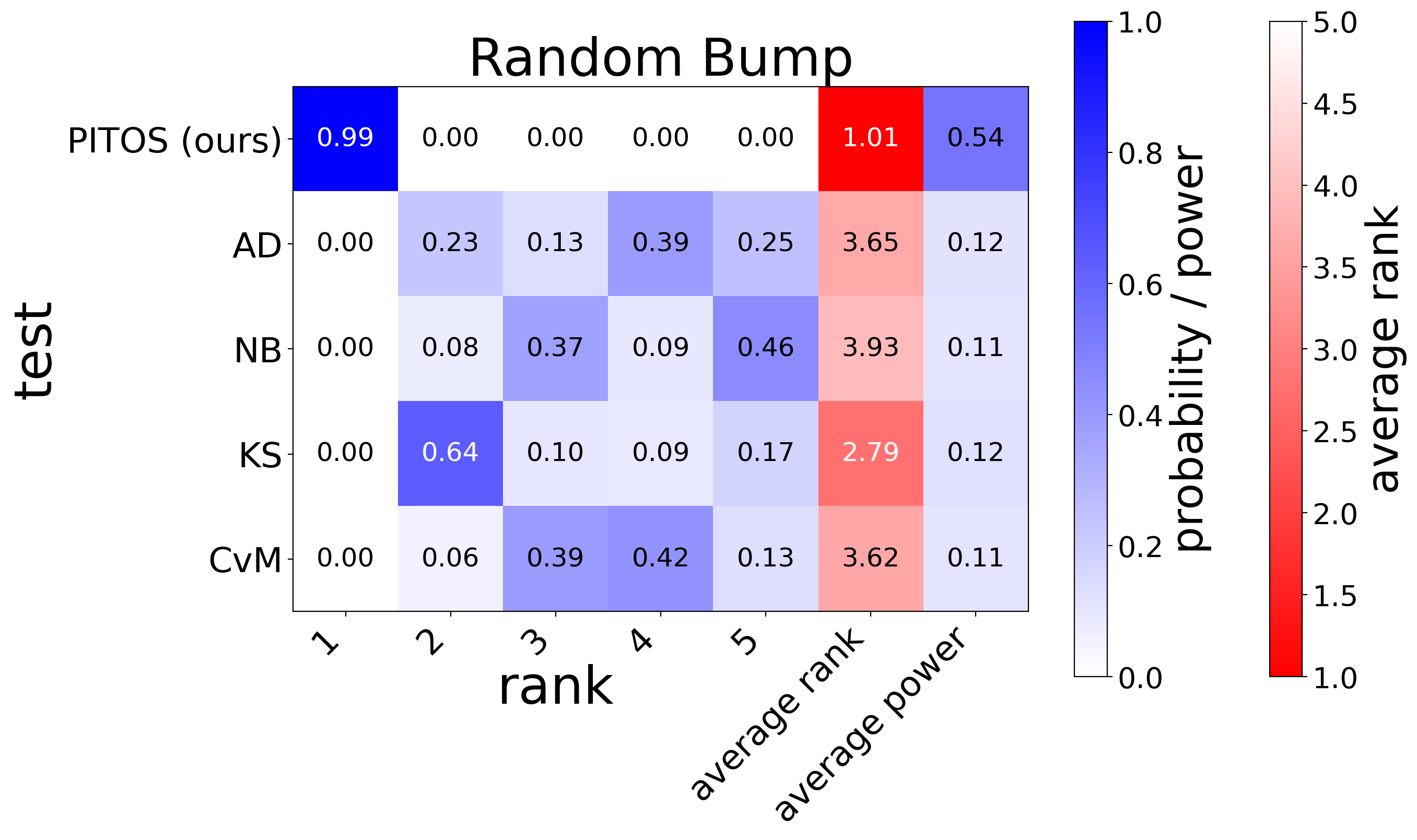}
    \hfill
    \includegraphics[width=0.48\textwidth]{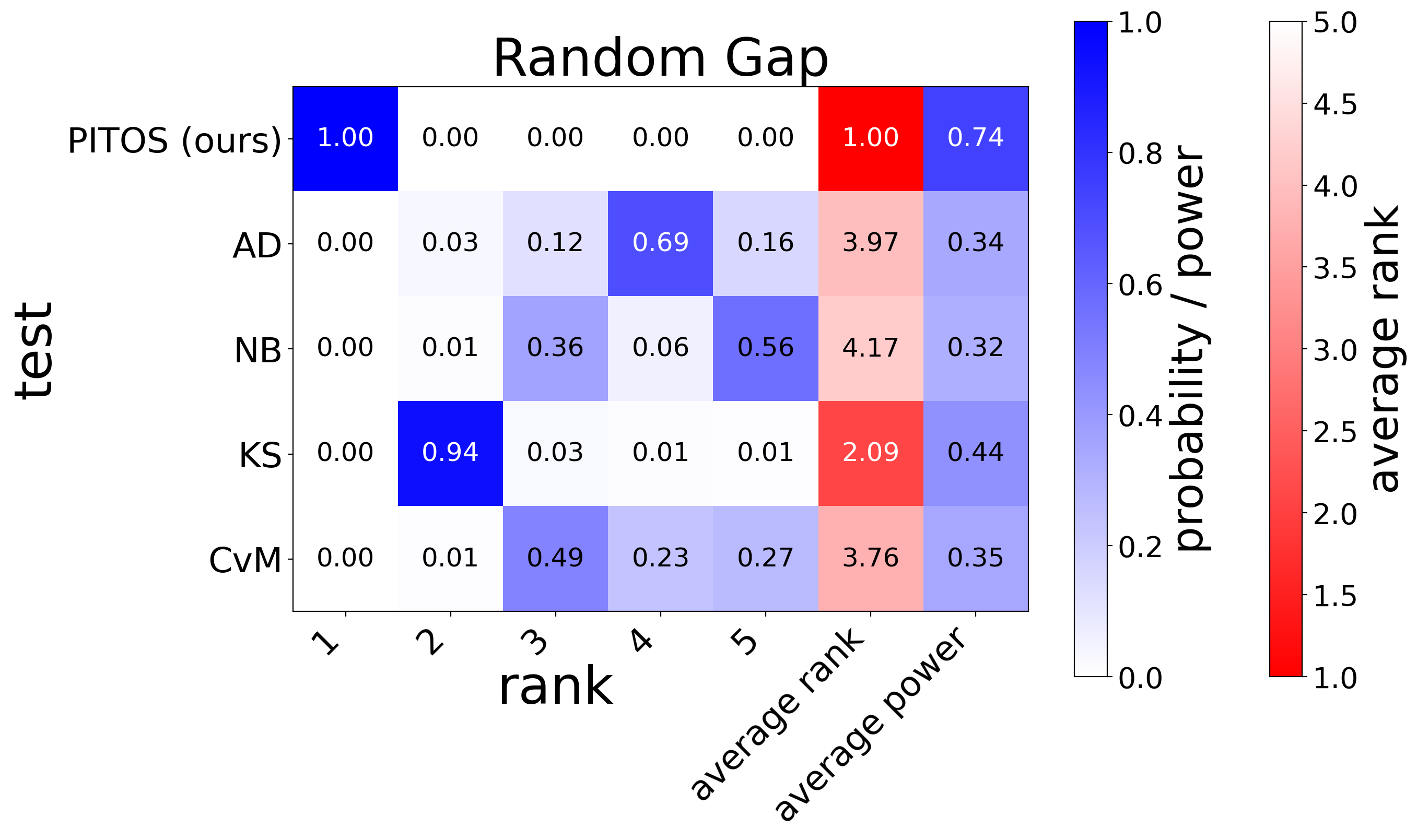}
    \hfill

    \caption{Power of GoF tests across families of data distributions.}
    \label{fig:power_families}
\end{figure}

Finally, we consider eight scenarios in which we randomly select a distribution $P_\theta$ from a given parametric family, and then approximate the power of each test on data from $P_\theta$ by simulating multiple i.i.d.\ data sets from $P_\theta$.
The scenarios considered are (1) Symmetric Heavy-Tailed, (2) Symmetric Light-tailed, (3) Asymmetric Heavy-tailed, (4) Asymmetric Light-tailed, (5) Outliers, (6) Nearly Uniform, (7) Random Bump, and (8) Random Gap; see \cref{sec:randomized_scenarios} for details. 
For each scenario, we randomly select 1{,}000 distributions and, for each distribution, we approximate the power of each test by simulating 1{,}000 i.i.d.\ data sets of size $n = 100$.
We summarize the results by aggregating in two ways. 
First, for each $P_\theta$, we rank the tests according to their power, and compute the proportion of times that test $t$ has rank $r$ across distributions $P_\theta$.
Second, we simply compute the average 
power of each test across distributions $P_\theta$.

\cref{fig:power_families} shows the results.
We see a similar pattern of performance as in \cref{fig:power_stark_failures,fig:regular}. 
In the Nearly Uniform scenario, PITOS performs worse than AD, NB, and CvM; however, all of the tests exhibit average power between $0.28$ and $0.34$, so the difference in performance is relatively small.
PITOS is usually outperformed by NB in the light-tailed scenarios, but it typically ranks first or second in these scenarios.
In all of the other scenarios, PITOS dominates the rankings, particularly in the Outliers, Random Bump, and Random Gap scenarios.

\section{Discussion}
\label{sec:discussion}

We have introduced the PITOS test, and demonstrated that it is competitive with existing GoF tests across a wide range of data distributions. 
Moreover, for distributions characterized by large local deviations from uniformity, PITOS is often substantially more powerful than existing methods.
While we recommend running PITOS using our proposed sequence of pairs based on a transformed Halton sequence, the test can be customized by using different sequences in order to target certain types of alternative distributions.
Optimizing the choice of sequence to maximize power for a given set of alternative distributions is a potentially interesting direction for future work.
A related direction for future work would be to establish theoretical results on the power of the PITOS test.

\section*{Acknowledgements}
C.T.C.\ was supported by National Institutes of Health (NIH) Training Grant T32CA09337. 
J.W.M.\ was supported in part by the National Cancer Institute of the NIH under award number R01CA240299. 
The content is solely the responsibility of the authors and does not necessarily represent the official views of the National Institutes of Health. 

\bibliographystyle{biometrika}
\bibliography{bib}

\begin{thebibliography}{14}
\expandafter\ifx\csname natexlab\endcsname\relax\def\natexlab#1{#1}\fi

\bibitem[{Anderson \& Darling(1952)}]{anderson1952asymptotic}
\textsc{Anderson, T.~W.} \& \textsc{Darling, D.~A.} (1952).
\newblock Asymptotic theory of certain ``goodness of fit'' criteria based on stochastic processes.
\newblock \textit{The Annals of Mathematical Statistics} , 193--212.

\bibitem[{Arnold et~al.(2008)Arnold, Balakrishnan \& Nagaraja}]{arnold2008first}
\textsc{Arnold, B.~C.}, \textsc{Balakrishnan, N.} \& \textsc{Nagaraja, H.~N.} (2008).
\newblock \textit{A First Course in Order Statistics}.
\newblock SIAM.

\bibitem[{Bezanson et~al.(2017)Bezanson, Edelman, Karpinski \& Shah}]{Julia-2017}
\textsc{Bezanson, J.}, \textsc{Edelman, A.}, \textsc{Karpinski, S.} \& \textsc{Shah, V.~B.} (2017).
\newblock Julia: A fresh approach to numerical computing.
\newblock \textit{SIAM {R}eview} \textbf{59}, 65--98.

\bibitem[{Brockwell(2007)}]{brockwell2007universal}
\textsc{Brockwell, A.} (2007).
\newblock {Universal residuals: A multivariate transformation}.
\newblock \textit{Statistics \& Probability Letters} \textbf{77}, 1473--1478.

\bibitem[{Cram{\'e}r(1928)}]{cramer1928composition}
\textsc{Cram{\'e}r, H.} (1928).
\newblock {On the composition of elementary errors: First paper: Mathematical deductions}.
\newblock \textit{Scandinavian Actuarial Journal} \textbf{1928}, 13--74.

\bibitem[{D'Agostino \& Stephens(1986)}]{dagostino1986goftechniques}
\textsc{D'Agostino, R.~B.} \& \textsc{Stephens, M.~A.} (1986).
\newblock \textit{Goodness-of-Fit Techniques}.
\newblock USA: Marcel Dekker, Inc.

\bibitem[{Halton(1964)}]{halton1964algorithm}
\textsc{Halton, J.~H.} (1964).
\newblock {Algorithm 247: Radical-inverse quasi-random point sequence}.
\newblock \textit{Communications of the ACM} \textbf{7}, 701--702.

\bibitem[{Kolmogorov(1933)}]{kolmogorov1933sulla}
\textsc{Kolmogorov, A.} (1933).
\newblock Sulla determinazione empirica di una legge didistribuzione.
\newblock \textit{Giorn Dell'inst Ital Degli Att} \textbf{4}, 89--91.

\bibitem[{Kornblith et~al.(2018)}]{HypothesisTests_jl}
\textsc{Kornblith, S.} et~al. (2018).
\newblock {HypothesisTests.jl: Hypothesis tests for Julia}.
\newblock \url{https://github.com/JuliaStats/HypothesisTests.jl}.
\newblock Version 0.11.6.

\bibitem[{Lehmann \& Romano(2005)}]{lehmann2005testing}
\textsc{Lehmann, E.~L.} \& \textsc{Romano, J.~P.} (2005).
\newblock \textit{Testing Statistical Hypotheses}.
\newblock Springer.

\bibitem[{Liu \& Xie(2020)}]{liu2020cauchy}
\textsc{Liu, Y.} \& \textsc{Xie, J.} (2020).
\newblock Cauchy combination test: a powerful test with analytic p-value calculation under arbitrary dependency structures.
\newblock \textit{Journal of the American Statistical Association} \textbf{115}, 393--402.

\bibitem[{Neyman(1937)}]{neyman1937smooth}
\textsc{Neyman, J.} (1937).
\newblock ``{Smooth} test'' for goodness of fit.
\newblock \textit{Scandinavian Actuarial Journal} \textbf{1937}, 149--199.

\bibitem[{Neyman \& Pearson(1933)}]{neyman1933ix}
\textsc{Neyman, J.} \& \textsc{Pearson, E.~S.} (1933).
\newblock {IX. On the problem of the most efficient tests of statistical hypotheses}.
\newblock \textit{Philosophical Transactions of the Royal Society of London. Series A, Containing Papers of a Mathematical or Physical Character} \textbf{231}, 289--337.

\bibitem[{Von~Mises(1936)}]{von1936wahrscheinlichkeit}
\textsc{Von~Mises, R.} (1936).
\newblock \textit{{Wahrscheinlichkeit Statistik und Wahrheit: Einf{\"u}hrung in die neue Wahrscheinlichkeitslehre und ihre Anwendung}}.
\newblock Springer.

\end{thebibliography}


\clearpage
\setcounter{page}{1}
\setcounter{section}{0}
\setcounter{table}{0}
\setcounter{figure}{0}
\setcounter{algorithm}{0}
\renewcommand{\theHsection}{SIsection.\arabic{section}}
\renewcommand{\theHtable}{SItable.\arabic{table}}
\renewcommand{\theHfigure}{SIfigure.\arabic{figure}}
\providecommand{\theHalgorithm}{SIalgorithm.\arabic{algorithm}}
\renewcommand{\thepage}{S\arabic{page}}  
\renewcommand{\thesection}{S\arabic{section}}   
\renewcommand{\thetable}{S\arabic{table}}   
\renewcommand{\thefigure}{S\arabic{figure}}
\renewcommand{\thealgorithm}{S\arabic{algorithm}}

\begin{center}
{\Large Supplementary Material for ``A powerful goodness-of-fit test using the probability integral transform of order statistics''}
\end{center}

\thispagestyle{empty}

\section{Standard Goodness-of-Fit Tests}
\label{sec:standard_gof_tests}

In \cref{supp:tab:tests}, we summarize the commonly used goodness-of-fit (GoF) tests that serve as benchmarks in our comparisons.
For each of these tests, the data $X_1,\ldots,X_n$ are assumed to be i.i.d.\ and the null hypothesis is that $X_1,\ldots,X_n\sim\mathrm{Uniform}(0,1)$.
In the Neyman--Barton test, $\pi_1(x)$ and $\pi_2(x)$ are the first two orthogonal Legendre polynomials on $[0,1]$.


\begin{table}[h!]
    \centering
    \caption{Common GoF Tests for $\mathrm{Uniform}(0,1)$ null hypothesis.}
    \label{supp:tab:tests}
    
    \begin{tabularx}{\textwidth}{l >{\raggedright\arraybackslash}X}
        \toprule
        \textbf{Test} & \textbf{Test statistic $T(X_1,\ldots,X_n)$} \\
        \midrule
        \begin{tabular}[t]{@{}l@{}}
            Anderson--Darling (AD)\\
            \citep{anderson1952asymptotic}
        \end{tabular}
        & 
        $ -n - n^{-1}\sum_{i=1}^n (2 i-1) ( \log X_{(i)} + \log(1-X_{(n-i+1)})) $ \\
        \midrule
        \begin{tabular}[t]{@{}l@{}}
            $2^{\text{nd}}$-order Neyman--Barton (NB)\\
            \citep{neyman1937smooth}
        \end{tabular}
        & 
        \begin{tabular}[t]{@{}l@{}}
            $\sum_{j=1}^{2} (n^{-1/2} \sum_{i=1}^n \pi_j(X_i))^2$ \\
            where $\pi_1(x) = 2 \sqrt{3} x$ and $\pi_2(x) = \sqrt{5}(6x^2 - 0.5)$
        \end{tabular} \\
        \midrule
        \begin{tabular}[t]{@{}l@{}}
            Kolmogorov--Smirnov (KS)\\
            \citep{kolmogorov1933sulla}
        \end{tabular}
        & 
        \begin{tabular}[t]{@{}l@{}}
            $ \sup_{t \in [0,1]} |F_n(t) - t| $ \\
            where $F_n(t) = \frac{1}{n} \sum_{i=1}^n \1{X_i \leq t}$
        \end{tabular} \\
        \midrule
        \begin{tabular}[t]{@{}l@{}}
            Cramér--von Mises (CvM)\\
            \citep{cramer1928composition,von1936wahrscheinlichkeit} 
        \end{tabular}
        & 
        \begin{tabular}[t]{@{}l@{}}
            $ 1/(12 n) + \sum_{i=1}^n \big(\frac{2 i - 1}{2 n} - X_{(i)}\big)^2 $ \\
        \end{tabular} \\
        \midrule
        \begin{tabular}[t]{@{}l@{}}
            Likelihood ratio test (LRT) \\
            \citep{neyman1933ix} 
        \end{tabular}
        & 
        \begin{tabular}[t]{@{}l@{}}
            $ \sum_{i=1}^n \log f_1(X_i)$ \\
            where $f_1(x)$ is the density of the alternative distribution
        \end{tabular} \\
        \bottomrule
    \end{tabularx}
\end{table}

For the Neyman-Barton, Cramér--von Mises, and likelihood ratio tests, we implement the test statistic in the Julia programming language~\citep{Julia-2017}
and compute p-values using empirical null distributions computed based on 100{,}000 simulated i.i.d.\ data sets from the $\mathrm{Uniform}(0,1)$ distribution, for any given $n$.
For Anderson--Darling and Kolmogorov--Smirnov, we use existing implementations from \textsf{HypothesisTests.jl} library in Julia \citep{HypothesisTests_jl}.

\section{Generalized Rosenblatt Transform}
\label{sec:generalized_rosenblatt}

For a random vector $Y_{1:n} = (Y_1, \ldots, Y_n)\in\mathbb{R}^n$, define the conditional CDFs $F_1, \ldots, F_n$ as
$F_1(y_1) = P(Y_1 \leq y_1)$ and 
$F_k(y_k \mid y_{1:(k-1)}) = P(Y_k \leq y_k \mid Y_{1:(k-1)} = y_{1:(k-1)})$ for $k \in \{2,3,\ldots,n\}$.
Likewise, define 
$F_1^{-}(y_1) = P(Y_1 < y_1)$ and
$F_k^{-}(y_k \mid y_{1:(k-1)}) = P(Y_k < y_k \mid Y_{1:(k-1)} = y_{1:(k-1)})$ for $k \in \{2,3,\ldots,n\}$.

For $y_1,\ldots,y_n\in\mathbb{R}$ and $u_1,\ldots,u_n \in(0,1)$, define the function $h(y_{1:n},u_{1:n})\in\mathbb{R}^n$ such that
\begin{align*}
h_1(y_{1:n},u_{1:n}) &= u_1 F_1(y_1) + (1-u_1) F_1^{-}(y_1) \\
h_k(y_{1:n},u_{1:n}) &= u_k F_k(y_k \mid y_{1:(k-1)}) + (1 - u_k) F_k^{-}(y_k \mid y_{1:(k-1)})
\end{align*}
for $k \in \{2,3,\ldots,n\}$.
If $U_1,\ldots,U_n$ i.i.d.\ $\sim\mathrm{Uniform}(0,1)$ and $X_{1:n} = h(Y_{1:n},U_{1:n})$ then $X_1, \ldots, X_n$ are i.i.d.\  
$\mathrm{Uniform}(0,1)$ random variables.
A proof is provided by~\citet{brockwell2007universal}.

\clearpage

\section{Randomly Generated Distributions in Simulation Study}
\label{sec:randomized_scenarios}

\begin{figure}[ht]
    \centering
    \includegraphics[width=0.23\textwidth]{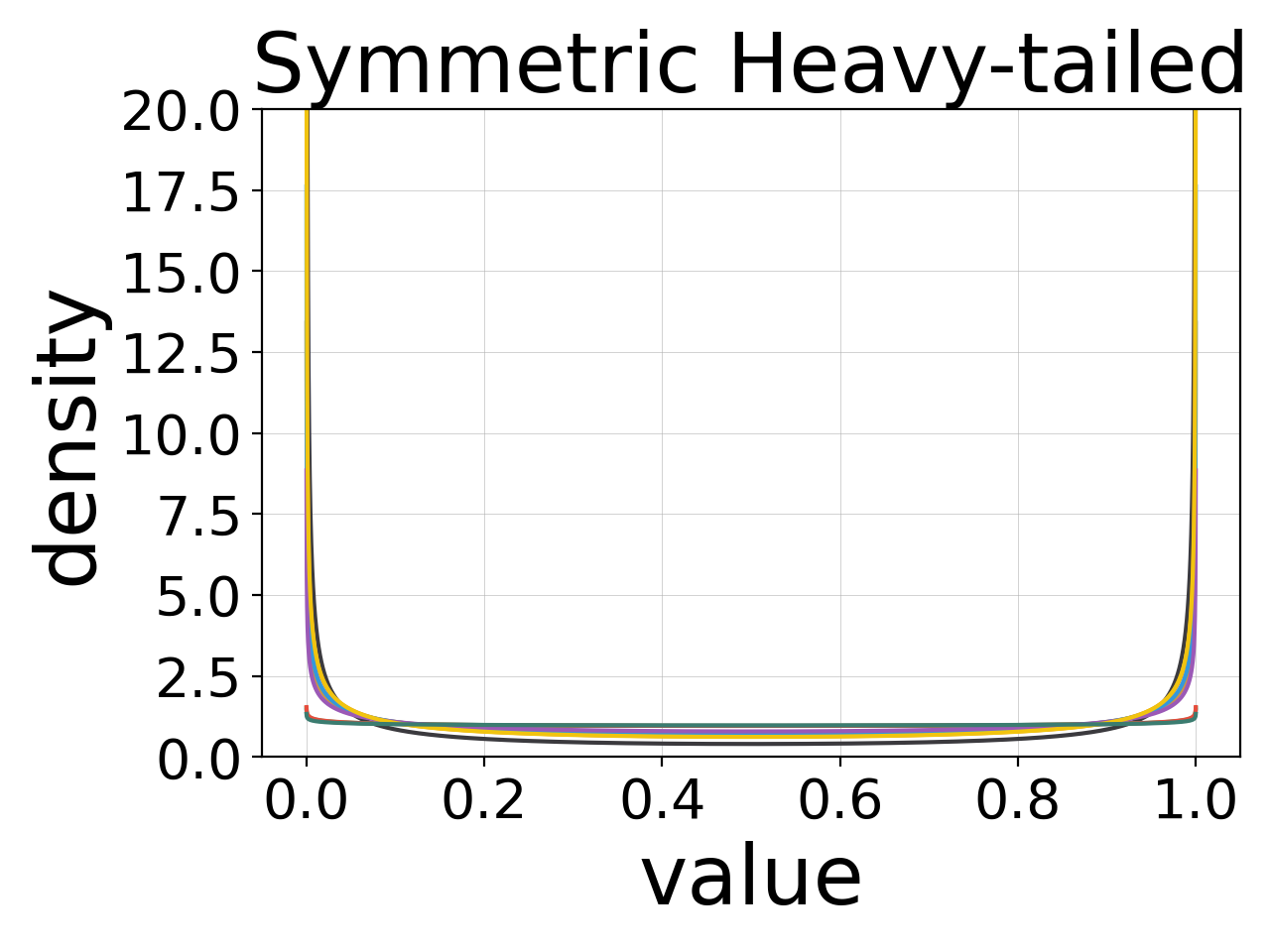}
    \hfill
    \includegraphics[width=0.23\textwidth]{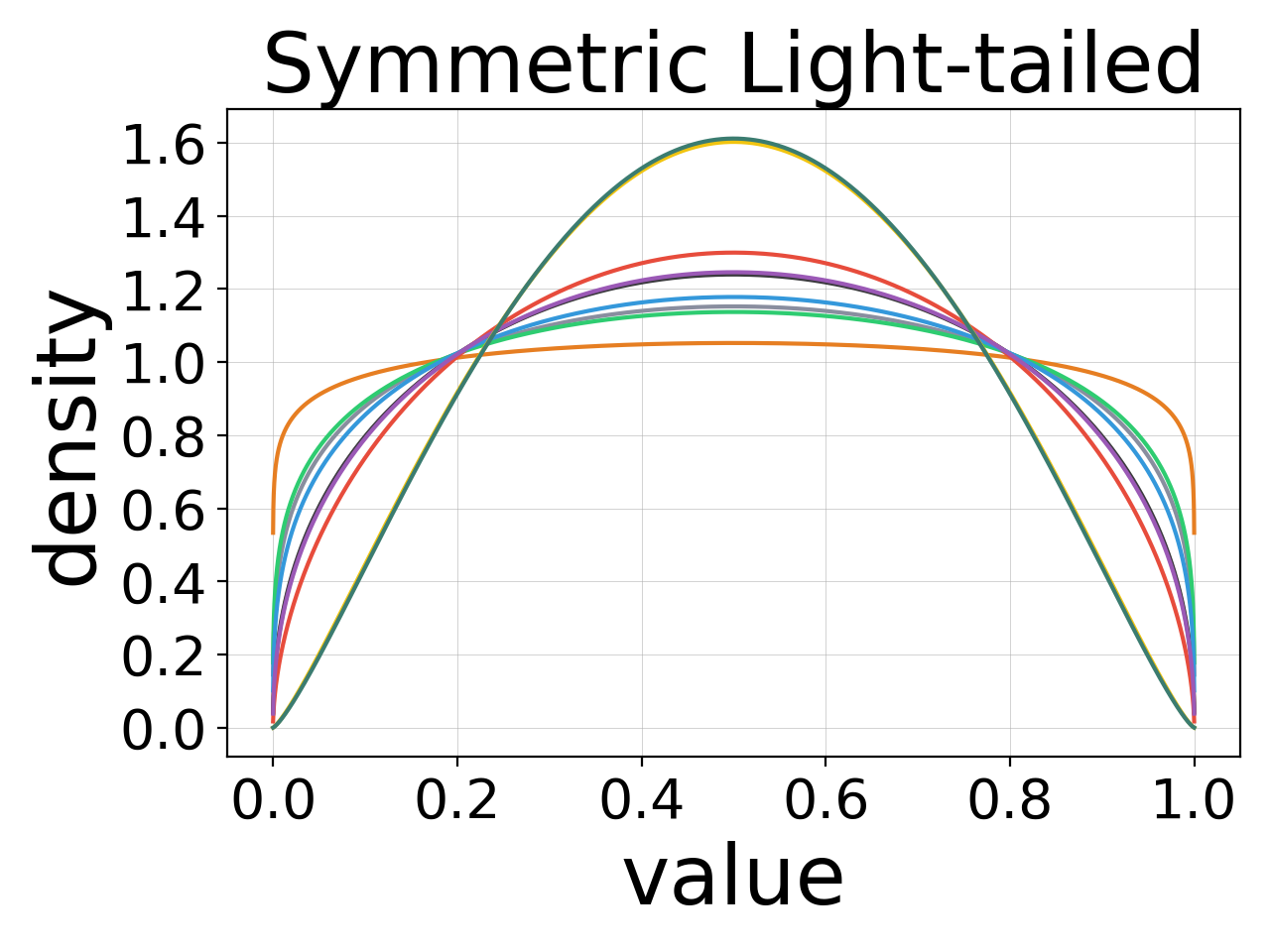}
    \hfill 
    \includegraphics[width=0.23\textwidth]{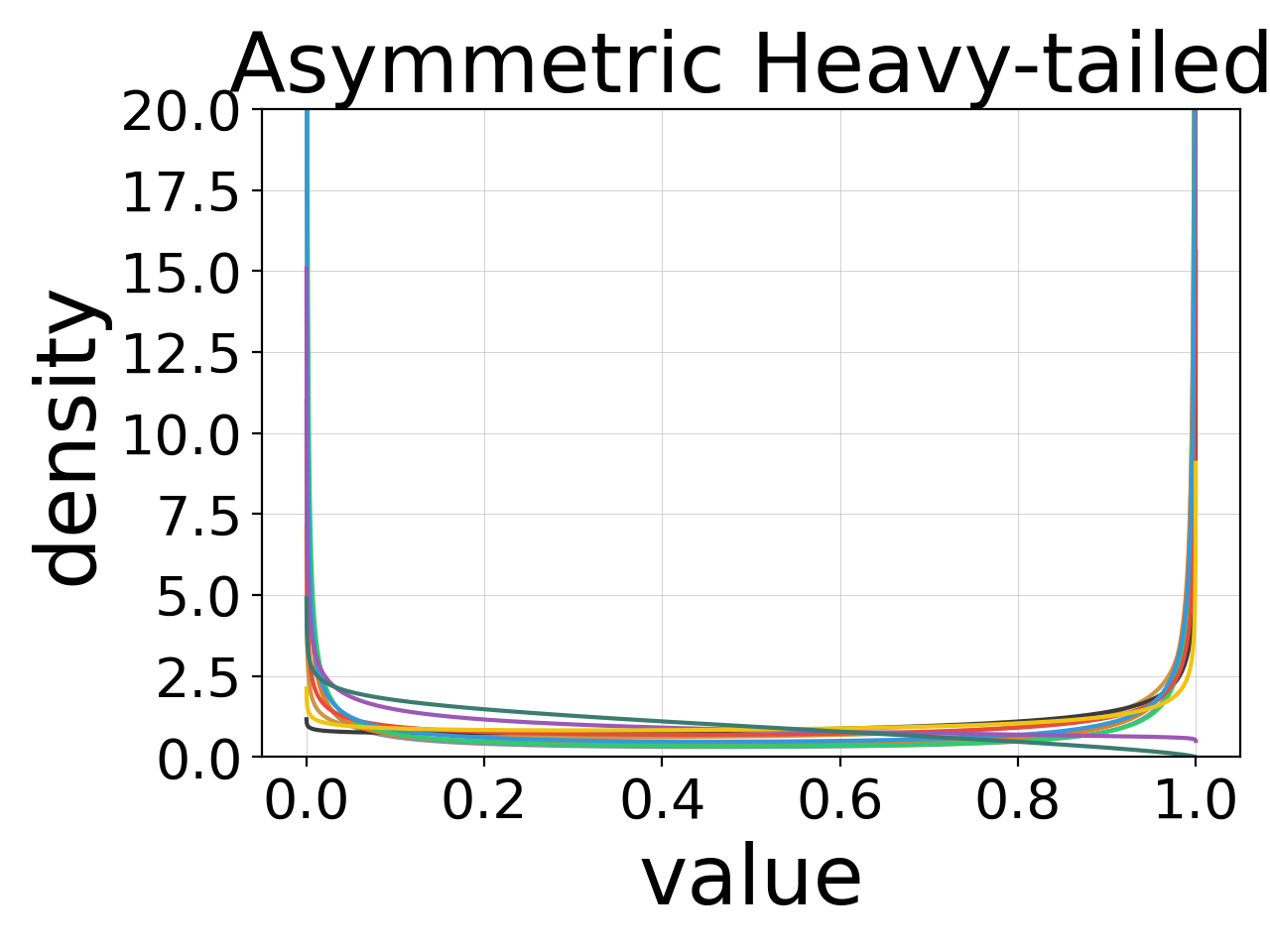}
    \hfill
    \includegraphics[width=0.23\textwidth]{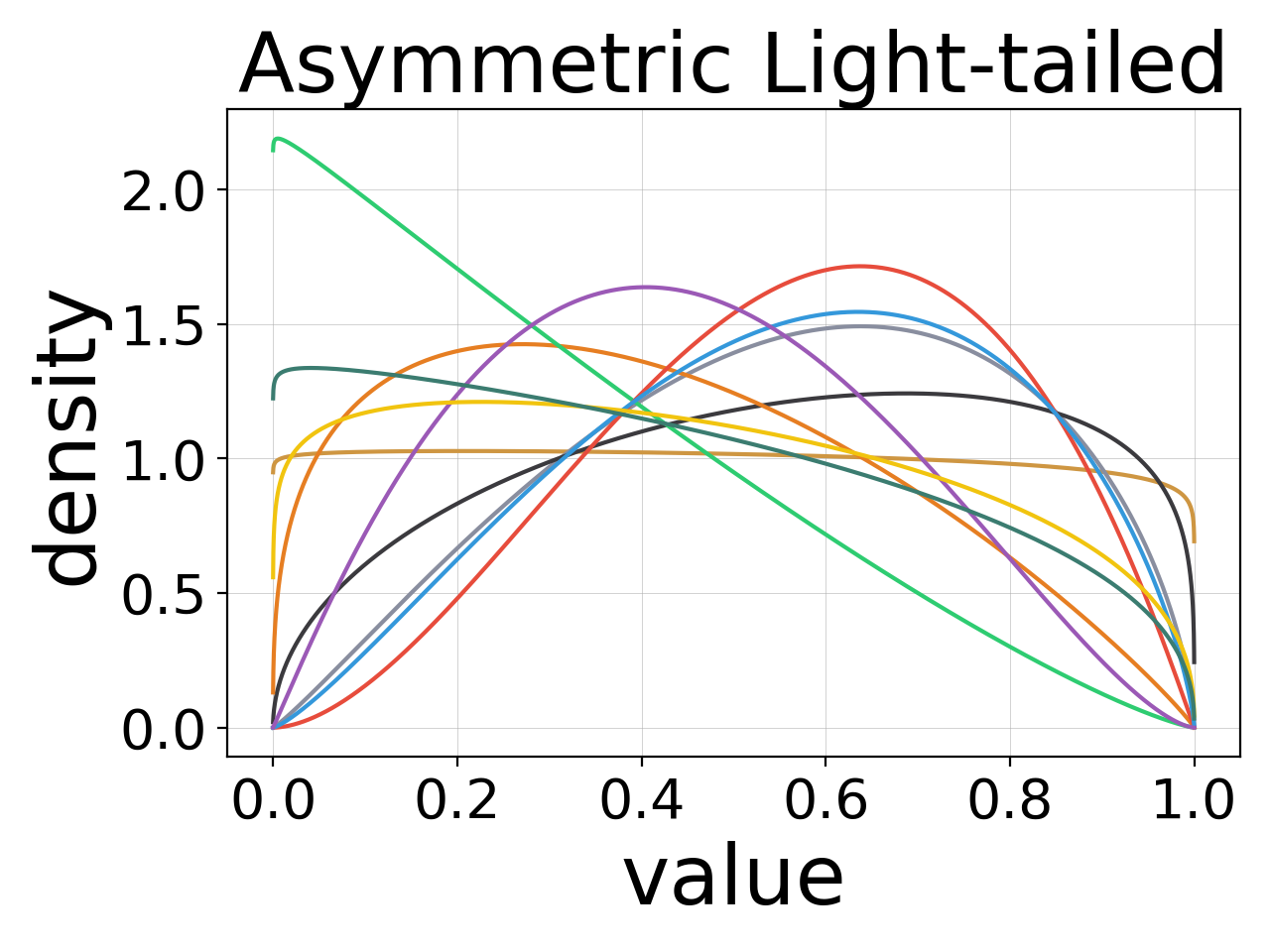}
    \hfill
    \includegraphics[width=0.23\textwidth]{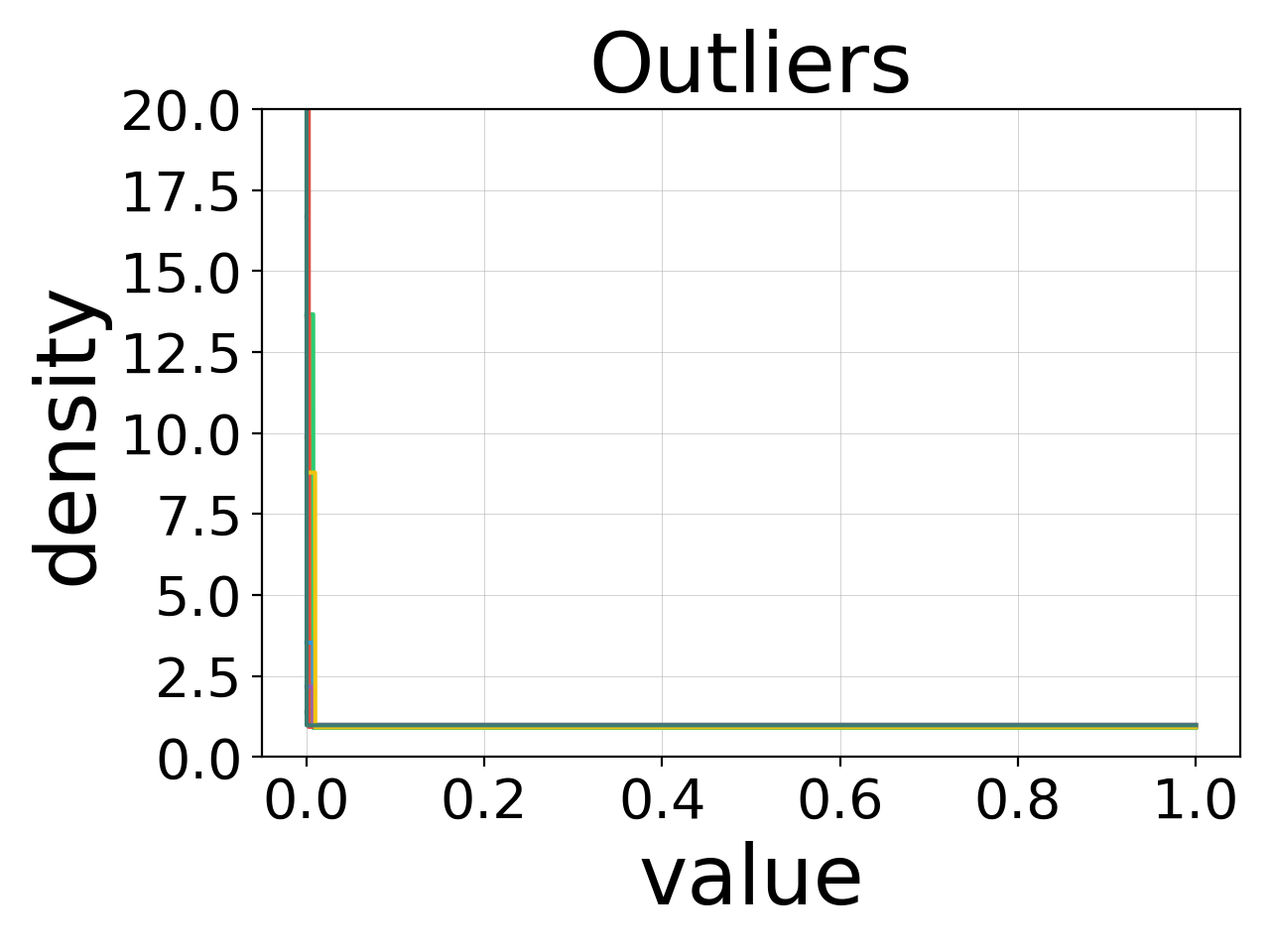}
    \hfill
    \includegraphics[width=0.23\textwidth]{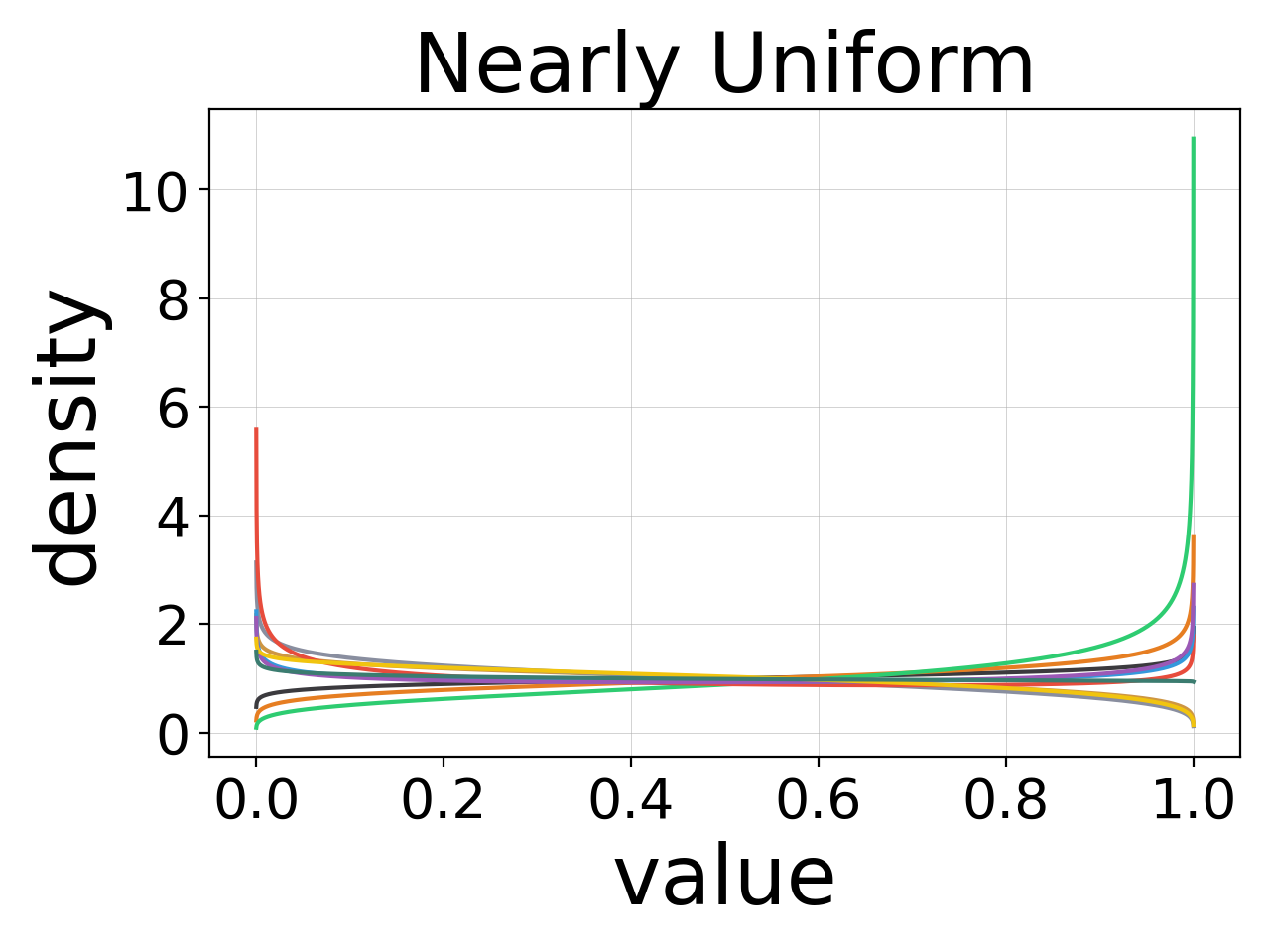}
    \hfill
    \includegraphics[width=0.23\textwidth]{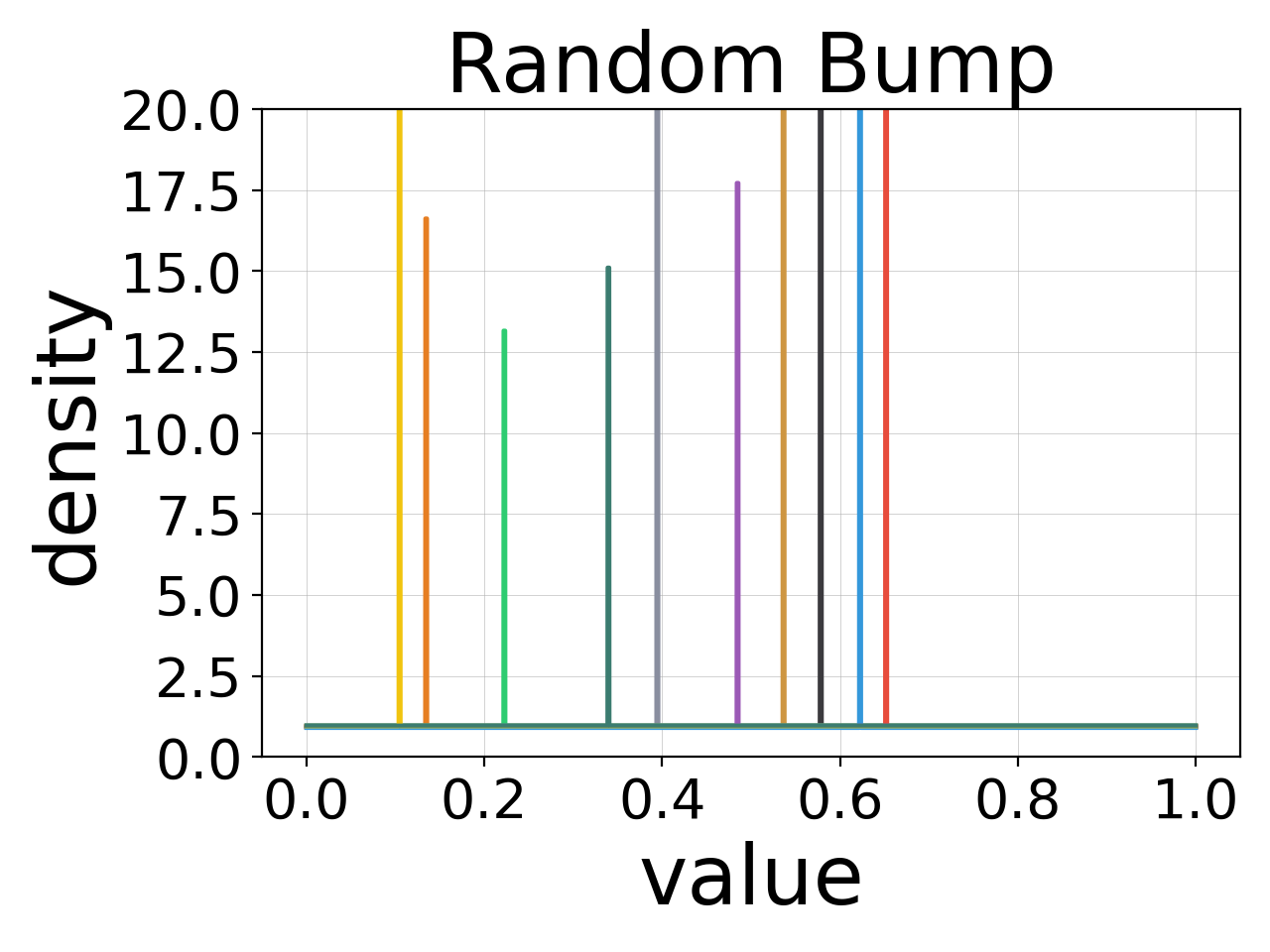}
    \hfill
    \includegraphics[width=0.23\textwidth]{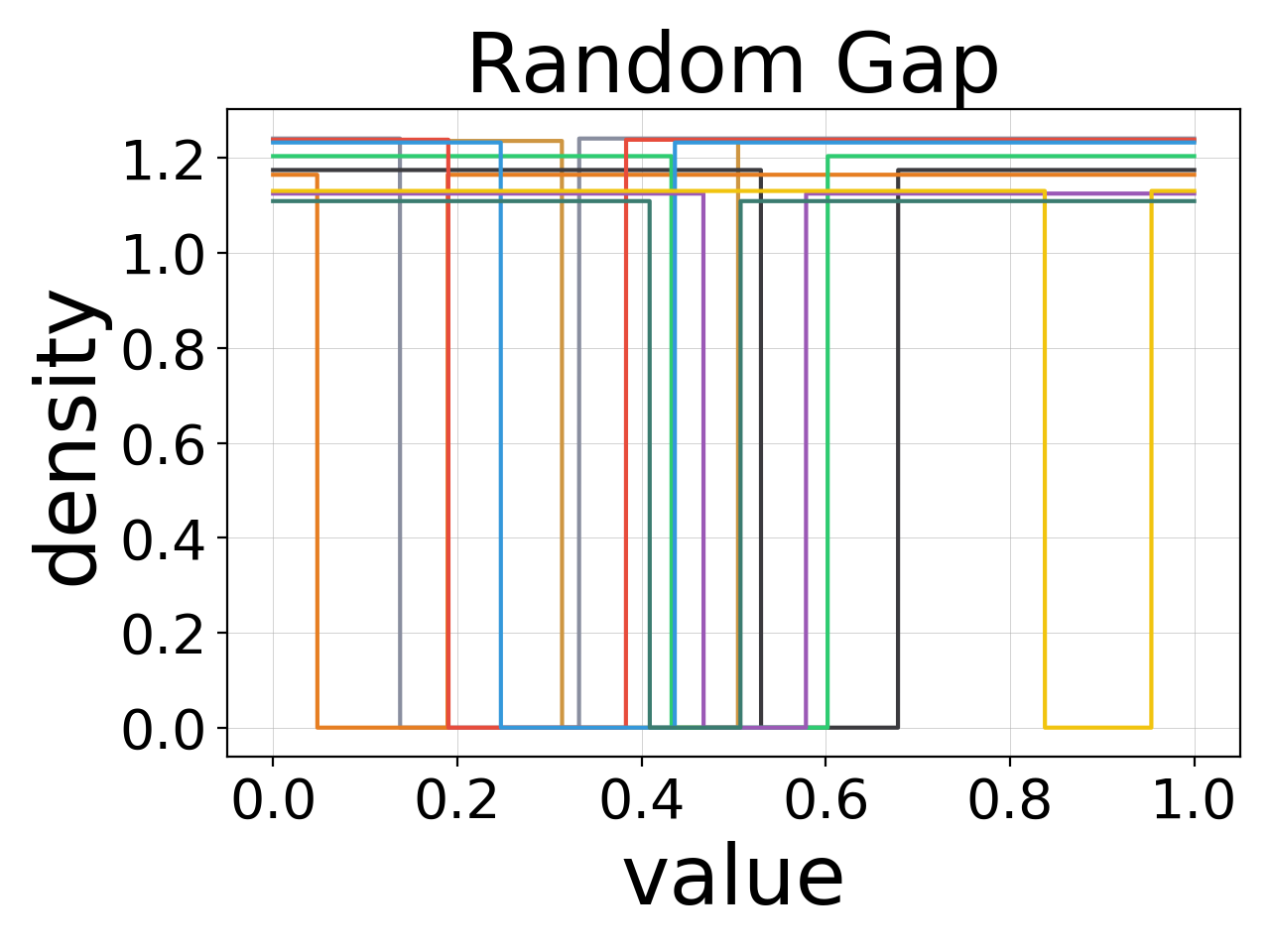}
    \caption{Examples of randomly sampled densities for each scenario in \cref{fig:power_families}.}
    \label{supp:fig:family_dists}
\end{figure}

\cref{supp:tab:family_dists} describes the sampling process used to generate randomized distributions for each scenario in \cref{fig:power_families}. 
Here, $\delta_x$ denotes the unit point mass at $x$, $U(a,b)$ denotes the uniform distribution on the interval $(a,b)$,
and $\mathrm{Gamma}(\alpha,\beta)$ denotes the Gamma distribution with density $f(x) = x^{\alpha-1} \exp(-x/\beta) / (\beta^\alpha\Gamma(\alpha))$ for $x>0$.
Visualizations of several randomly sampled densities for each scenario are shown in~\cref{supp:fig:family_dists}.

\begin{table}[ht]
    \centering
    \caption{Procedures for randomly sampling distributions for each scenario in \cref{fig:power_families}.}
    \label{supp:tab:family_dists}
    \renewcommand{\arraystretch}{1.2}
    \resizebox{\textwidth}{!}{
    \begin{tabular}{@{}c c l@{}}
    \toprule
    Scenario & Parametric family & Random generation of parameters \\
    \midrule
    Symmetric Heavy-tailed 
        & $\mathrm{Beta}(\mu \sigma, (1-\mu)\sigma)$ 
        & \begin{tabular}[t]{@{}l@{}}
            $\mu \sim \delta_{1/2},\, \sigma \sim \mathrm{Gamma}(3, 1/2)$ \\
            Draw $(\mu,\sigma)$ until $\min(\mu\sigma,(1-\mu)\sigma) \leq 1$
          \end{tabular} \\
    \midrule
    Symmetric Light-tailed 
        & $\mathrm{Beta}(\mu \sigma,\, (1-\mu)\sigma)$ 
        & \begin{tabular}[t]{@{}l@{}}
            $\mu \sim \delta_{1/2},\, \sigma \sim \mathrm{Gamma}(5, 1/2)$ \\
            Draw $(\mu,\sigma)$ until $\min(\mu\sigma,(1-\mu)\sigma) > 1$
          \end{tabular} \\
    \midrule
    Asymmetric Heavy-tailed 
        & $\mathrm{Beta}(\mu \sigma,\, (1-\mu)\sigma)$ 
        & \begin{tabular}[t]{@{}l@{}}
            $\mu \sim \mathrm{Beta}(2,2),\, \sigma \sim \mathrm{Gamma}(3, 1/2)$ \\
            Draw $(\mu,\sigma)$ until $\min(\mu\sigma,(1-\mu)\sigma) \leq 1$
          \end{tabular} \\
    \midrule
    Asymmetric Light-tailed 
        & $\mathrm{Beta}(\mu \sigma,\, (1-\mu)\sigma)$ 
        & \begin{tabular}[t]{@{}l@{}}
            $\mu \sim \mathrm{Beta}(2,2),\, \sigma \sim \mathrm{Gamma}(5, 1/2)$ \\
            Draw $(\mu,\sigma)$ until $\min(\mu\sigma,(1-\mu)\sigma) > 1$
          \end{tabular} \\
    \midrule
    Outliers 
        & $\pi \, U(0, b) + (1-\pi) \, U(0, 1)$
        & $\pi \sim U(0, 0.1),\, b \sim U(0, 0.01)$ \\
    \midrule
    Nearly Uniform 
        & $\mathrm{Beta}(\mu \sigma,\, (1-\mu)\sigma)$ 
        & $\mu \sim \mathrm{Beta}(50,50),\, \sigma \sim \mathrm{Gamma}(100, 1/50)$ \\
    \midrule 
    Random Bump
        & $\pi \, U(m - 0.001, m + 0.001) + (1-\pi) \, U(0, 1)$
        & $m \sim U(0.001, 0.999),\, \pi \sim U(0, 0.1)$ \\
    \midrule
    Random Gap
        & $\Big(\frac{m - w}{1-2 w}\Big) \, U(0, m - w) + \Big(\frac{1 - (m + w)}{1-2 w}\Big) \, U(m + w,1)$
        & $m \sim U(0.1, 0.9),\, w \sim U(0.025, 0.1)$ \\
    \bottomrule
    \end{tabular}
    }
\end{table}

\clearpage
\section{Algorithms for Implementing the PITOS Goodness-of-Fit Test}
\label{sec:algorithms}

In this section, we provide step-by-step algorithms for computing the PITOS p-value $p^*$ (\cref{alg:pitos}), generating Halton sequences (\cref{alg:1D_Halton}), and our proposed method of generating the sequence of pairs to be used in the PITOS algorithm (\cref{supp:alg:tail_weighted_halton}).

\subsection{Main PITOS algorithm}

The algorithm takes $O(n \log n)$ time for any collection of pairs $\mathcal{I}$ of size $O(n \log n)$.
We write $G(x,a,b)$ to denote the CDF of $\mathrm{Beta}(a,b)$ at $x$, that is, $G(x,a,b) = \int_0^x \frac{1}{\mathrm{B}(a,b)}t^{a-1}(1-t)^{b-1} d t$.

\begin{algorithm}
\caption{~~PITOS goodness-of-fit test for $\mathrm{Uniform}(0,1)$ null distribution}
\label{alg:pitos}
\begin{algorithmic}
\Require{Data $x_1,\ldots,x_n \in [0,1]$ and pairs $\mathcal{I}=((i_1,j_1),\ldots,(i_m,j_m))$ where $i_k,j_k\in\{1,\ldots,n\}$}
\Ensure{Approximate p-value $p^*$}
\begin{enumerate}
    \item Sort $x_1,\ldots,x_n$ to obtain the order statistics $x_{(1)} \leq \cdots \leq x_{(n)}$
    \item \textbf{for} $(i,j) \in \mathcal{I}$ \textbf{do}
        \begin{enumerate}
            \item[] \textbf{if} $i = j$ \textbf{then}
                \begin{itemize}
                \item[] $u_{ij} \gets G(x_{(j)}, j, n - j + 1)$
                \end{itemize}
            \item[] \textbf{elseif} $i < j$ \textbf{then}
                \begin{itemize}
                \item[] $u_{ij} \gets G((x_{(j)} - x_{(i)})/(1 - x_{(i)}), j - i, n - j + 1)$
                \end{itemize}
            \item[] \textbf{else}
                \begin{itemize}
                \item[] $u_{ij} \gets G(x_{(j)}/x_{(i)}, j, i - j)$
                \end{itemize}
            \item[] \textbf{end}
        \item[] $p_{ij} \gets 2 \min(u_{ij}, 1 - u_{ij})$
        \end{enumerate}
        \textbf{end}
    \item\label{step:cauchy_combo} $p \gets 1 - F_{\mathrm{Cauchy}} \Big( \frac{1}{m} \sum_{(i,j) \in \mathcal{I}} F^{-1}_{\mathrm{Cauchy}}(1-p_{ij}) \Big)$
    \item $p^* \gets \min(1, 1.15 p)$
    \item \textbf{return} $p^*$
\end{enumerate}
\end{algorithmic}
\end{algorithm}

Since a given pair $(i,j)$ may occur more than once in $\mathcal{I}$, this algorithm could potentially be sped up further by caching the value of $p_{i j}$ for each $(i,j)$ as it is computed, and re-using the previously computed value of $p_{i j}$ instead of re-computing it if $(i,j)$ is encountered again.

\newpage
\subsection{Generating PITOS pairs using a discretized transformed Halton sequence}

For completeness, we next provide a procedure for generating Halton sequences \citep{halton1964algorithm} in \cref{alg:1D_Halton}, which is used as a subroutine in \cref{supp:alg:tail_weighted_halton}.
In \cref{alg:1D_Halton}, we write $i~\mathrm{mod}~b$ to denote the remainder when dividing $i$ by $b$, where $i$ and $b$ are positive integers.

\begin{figure}
    \centering
    \includegraphics[width=0.48\textwidth]{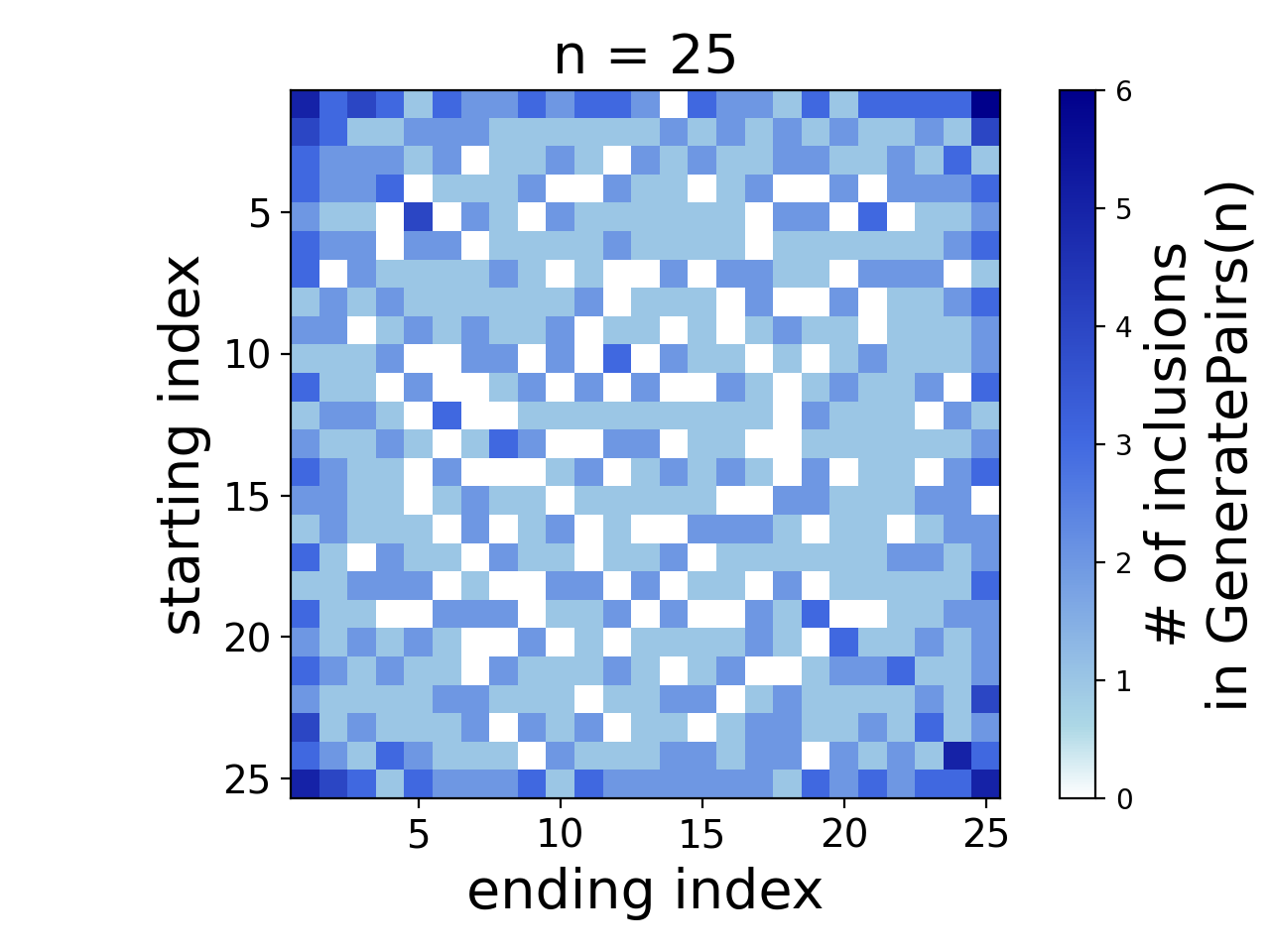}
    \hfill 
    \includegraphics[width=0.48\textwidth]{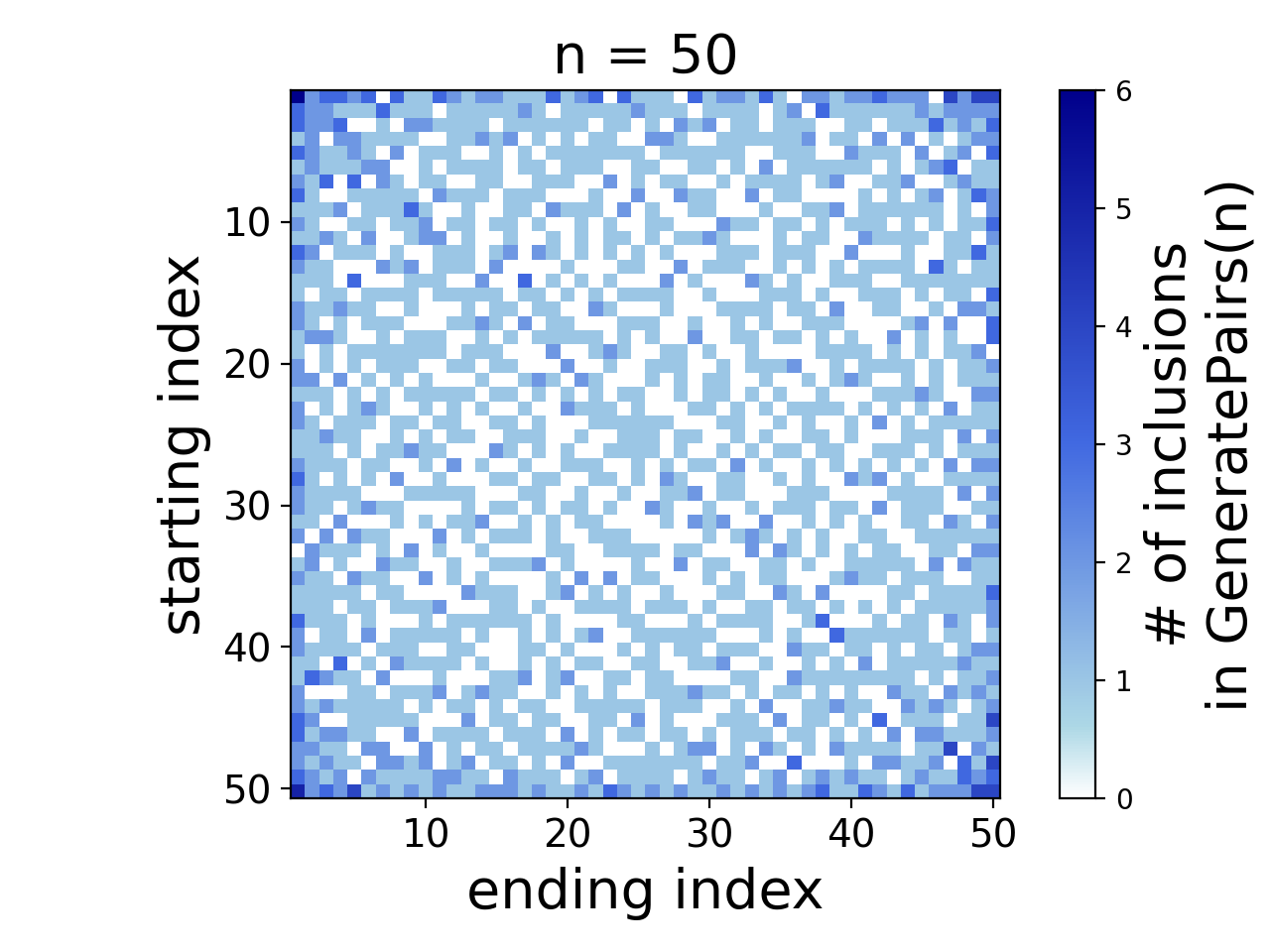}
    \hfill 
    \includegraphics[width=0.48\textwidth]{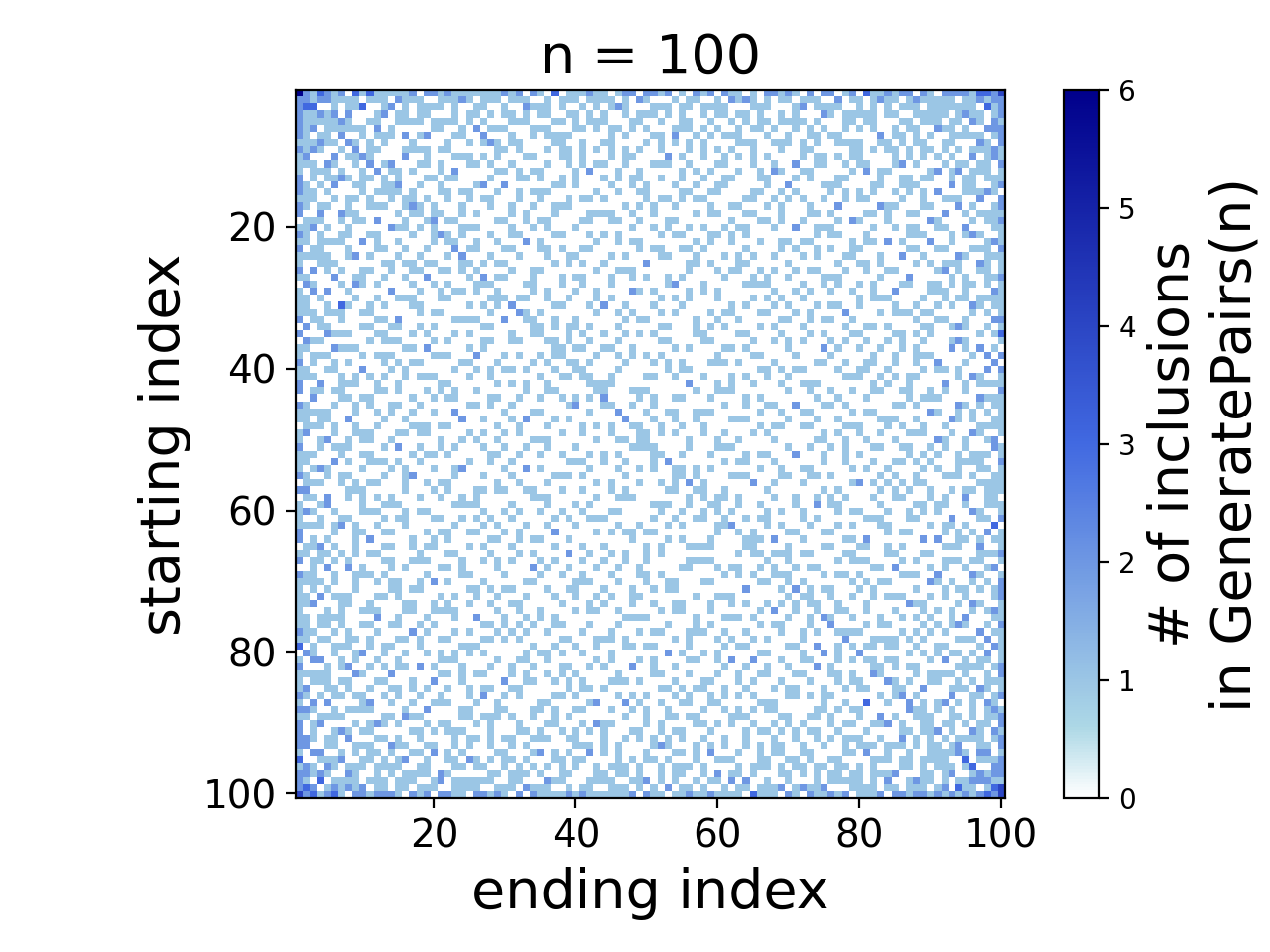}
    \hfill 
    \includegraphics[width=0.48\textwidth]{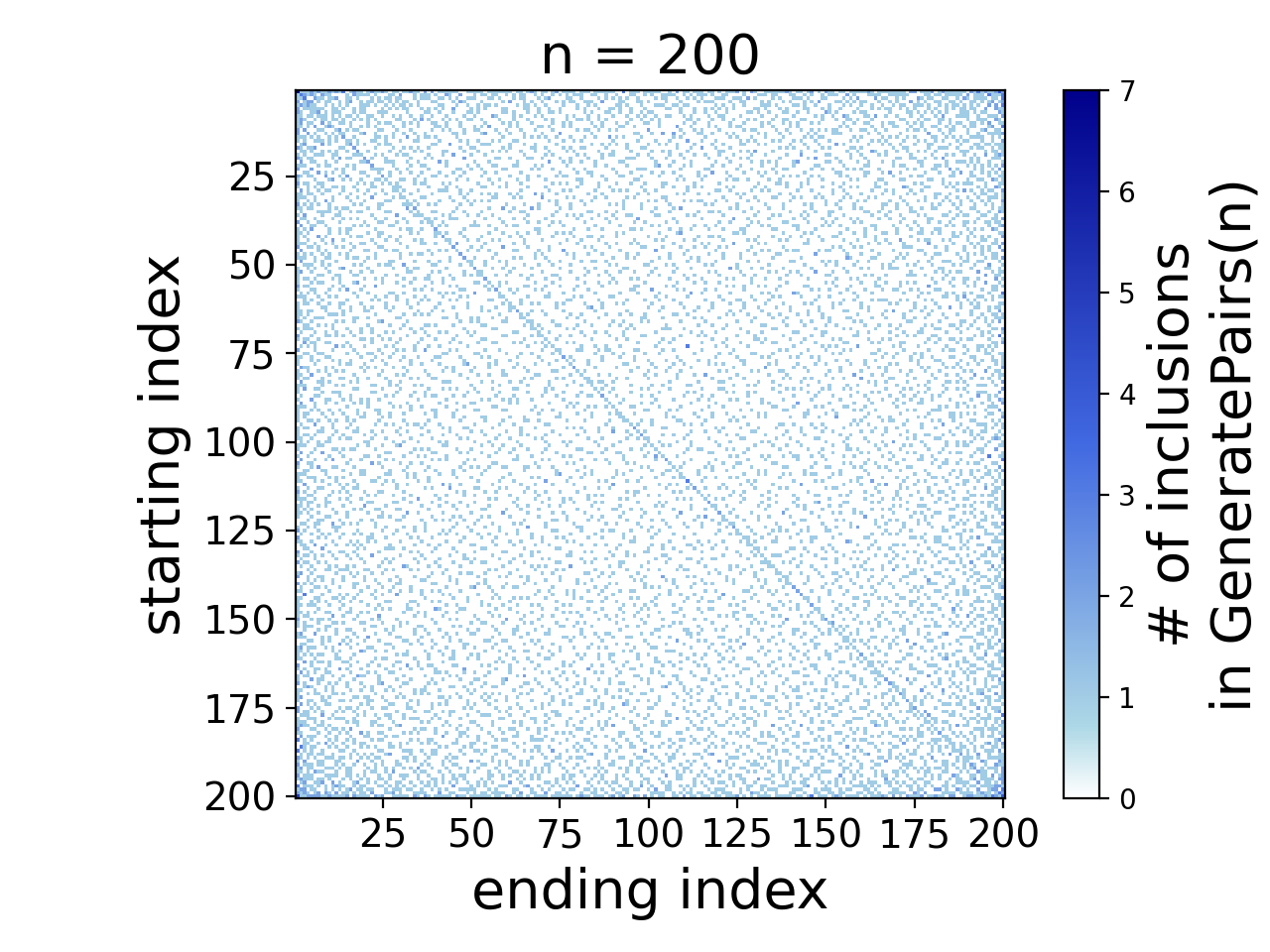}
    \caption{Number of times each pair of indices $(i,j)$ is included in \texttt{GeneratePairs(n)}.}
    \label{supp:fig:tail_weighted_halton_heatmap}
\end{figure}

\cref{supp:alg:tail_weighted_halton} details how we compute the sequence $\mathcal{I}$ to be used in \cref{alg:pitos}.
To visualize the sequence of pairs $\mathcal{I} = ((i_1,j_1),\ldots,(i_m,j_m))$ produced by \cref{supp:alg:tail_weighted_halton}, we plot heatmaps in  \cref{supp:fig:tail_weighted_halton_heatmap} showing the number of times each $(i,j)$ is included in $\mathcal{I}$, for a range of sample sizes $n\in\{25,50,100,200\}$.

\begin{algorithm}
\caption{~~\texttt{Halton}: Generate one-dimensional element of Halton sequence}
\label{alg:1D_Halton}
\begin{algorithmic}
\Require{Index $i\in\{1,2,\ldots\}$ and base $b\in\{1,2,\ldots\}$}
\Ensure{The $i$th element of the one-dimensional Halton sequence with base $b$}
\begin{enumerate}
\item $f\gets 1$
\item $x\gets 0$
\item \textbf{while} $i > 0$ \textbf{do}
    \begin{itemize}
    \item[] $f \gets f / b$
    \item[] $x \gets x + (i~\mathrm{mod}~b) \cdot f$
    \item[] $i \gets \lfloor i/b \rfloor$
    \end{itemize}
\item[] \textbf{end}
\item \textbf{return} $x$
\end{enumerate}
\end{algorithmic}
\end{algorithm}

\begin{algorithm}
\caption{~~\texttt{GeneratePairs}: Generate sequence of pairs for PITOS test in \cref{alg:pitos}}
\label{supp:alg:tail_weighted_halton}
\begin{algorithmic}
\Require{Sample size $n \in \{1,2,\ldots\}$}
\Ensure{Sequence of pairs $\mathcal{I} = ((i_1,j_1),\ldots,(i_m,j_m))$ for input to PITOS test}
\begin{enumerate}
\item $m \gets \lceil 10\, n \log n\rceil + n$
\item \textbf{for} $k = 1,\ldots,m-n$ \textbf{do}
    \begin{itemize}
    \item[] $u \gets \texttt{Halton}(k,2)$
    \item[] $v \gets \texttt{Halton}(k,3)$
    \item[] $x \gets F_{\text{Beta}(0.7, 0.7)}^{-1}(u)$
    \item[] $y \gets F_{\text{Beta}(0.7, 0.7)}^{-1}(v)$
    \item[] $i_k \gets \lceil n x \rceil$
    \item[] $j_k \gets \lceil n y \rceil$
    \end{itemize}
\item[] \textbf{end}
\item \textbf{for} $r = 1,\ldots,n$ \textbf{do}
    \begin{itemize}
        \item[] $i_{m-n+r} \gets r$
        \item[] $j_{m-n+r} \gets r$
    \end{itemize}
\item[] \textbf{end}
\item \textbf{return} $((i_1,j_1),\ldots,(i_m,j_m))$
\end{enumerate}
\end{algorithmic}
\end{algorithm}

\clearpage
\section{Approximately correct control of Type I error}

\begin{figure}[ht]
\centering
    \includegraphics[width=0.75\textwidth]{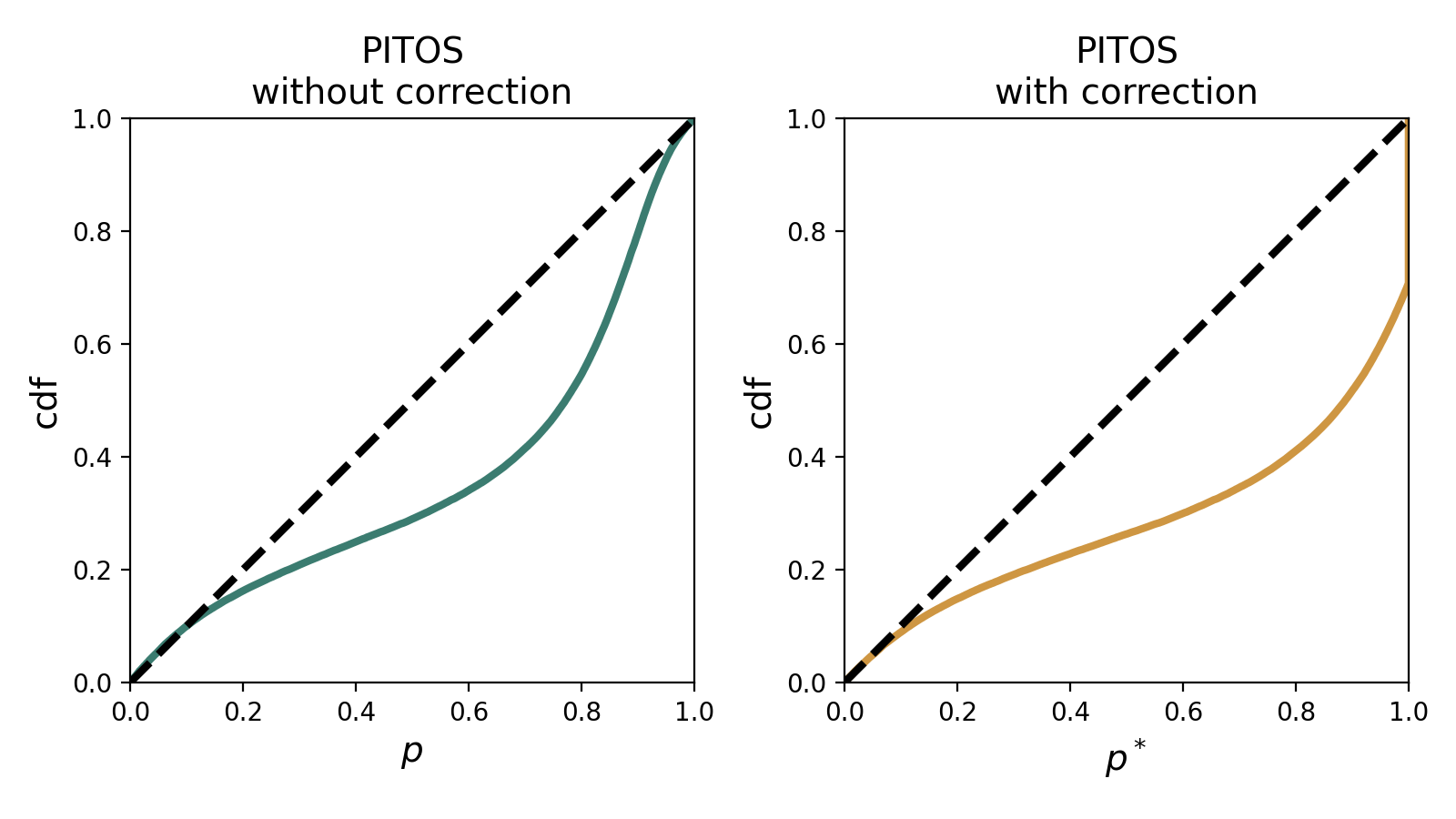} %
    \includegraphics[width=0.75\textwidth]{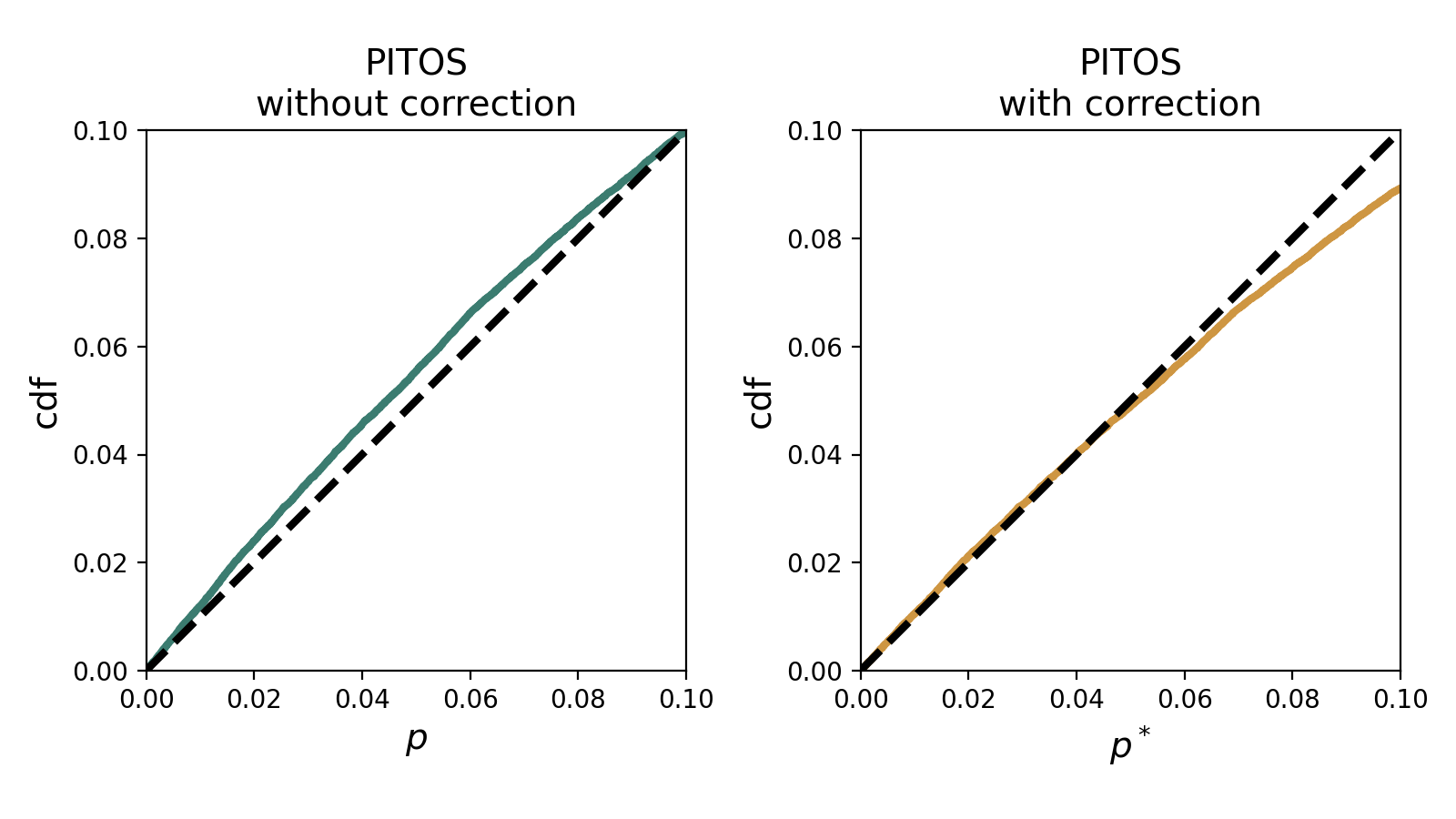}
    \caption{CDFs of PITOS p-values with and without the $1.15\times$ correction,  under the null hypothesis, for a sample size of $n = 30$. 
    Top plots show CDF over the full support of $[0,1]$.
    Bottom plots are zoomed in to $[0, 0.1]$; in practice, it is common to reject the null at levels $\alpha\in(0, 0.05]$. 
    }
    \label{supp:fig:PITOS_nulls}
\end{figure}

\cref{supp:fig:PITOS_nulls} shows the CDF of PITOS p-values with and without the $1.15\times$ correction (that is, $p^*$ and $p$, respectively), for i.i.d.\ data generated under the null hypothesis of $X_1,\ldots,X_n\sim \mathrm{Uniform}(0,1)$ with $n = 30$.
A p-value CDF above the $\mathrm{Uniform}(0,1)$ CDF (shown as a black dashed line) is too liberal, meaning that Type I error is larger than $\alpha$ when rejecting the null at level $\alpha$.
Meanwhile, a p-value CDF below the $\mathrm{Uniform}(0,1)$ CDF is conservative, meaning that Type I error is smaller than $\alpha$ when rejecting at level $\alpha$.

While the CDFs of both $p$ and $p^*$ are overly conservative for values greater than $0.2$, they are fairly close to the $\mathrm{Uniform}(0,1)$ CDF for values between $0$ and $0.1$, which is the range of values that matters in practice.
In the range $[0,0.01]$, the CDF of the uncorrected p-values $p$ is slightly liberal.
Meanwhile, the CDF of the corrected p-values $p^*$ is very close to the $\mathrm{Uniform}(0,1)$ CDF between $0$ and $0.05$, and is slightly conservative between $0.05$ and $0.1$.

\end{document}